\documentclass[onecolumn,
superscriptaddress,
rmp
]{revtex4-2}

\usepackage[x11names, rgb]{xcolor}

\definecolor{red}{HTML}{f54b1a}
\definecolor{pink}{HTML}{d19eb1}
\definecolor{orange}{HTML}{d3772e}
\definecolor{yellow}{HTML}{ebe85d}
\definecolor{green}{HTML}{0f6852}
\definecolor{lightblue}{HTML}{01abe9}
\definecolor{darkblue}{HTML}{1b346c}
\definecolor{tan}{HTML}{e5c39e}
\definecolor{darktan}{HTML}{af9e73}
\definecolor{grey}{HTML}{c3ced0}
\definecolor{darkgrey}{HTML}{9dadc4}
\definecolor{black}{HTML}{110d1b}
\definecolor{white}{HTML}{f1f8f1}

\usepackage{hyperref}
\hypersetup{
  colorlinks = true,
  linkcolor = red,
  citecolor = darkblue,
  urlcolor = lightblue
}

\usepackage{graphicx}

\usepackage{amsfonts,amssymb, amsmath}

\usepackage{float}
\makeatletter
\let\newfloat\newfloat@ltx
\makeatother

\usepackage{algorithm}
\usepackage[noend]{algpseudocode}
\algrenewcommand{\algorithmiccomment}[1]{$\vartriangleright$ #1}
\algrenewcommand{\algorithmicreturn}{\textbf{Return: }}
\algnewcommand\algorithmicinput{\textbf{Input: }}
\algnewcommand\Input{\State \algorithmicinput}

\def\eps{\varepsilon}

\def\RR{\mathbb{R}}  \def\ZZ{\mathbb{Z}}
\def\EE{\mathbb{E}}\def\PP{\mathbb{P}}

\def\<{\langle} \def\>{\rangle}

\DeclareRobustCommand{\ran}{\operatorname*{ran}}

\usepackage[utf8]{inputenc}
\usepackage[T1]{fontenc}
\usepackage{microtype}

\begin{document}

\title{Basic introduction to Freidlin--Wentzell theory}            

\author{Jiawei Yan}
\email{yann.jiawei@gmail.com}
\affiliation{Independent researcher, Shanghai 200135, China}

\begin{abstract}
Freidlin--Wentzell theory provides the quantitative foundation for rare, noise-induced transitions in metastable systems, predicting optimal transition paths, the exponential scaling of mean escape times, and the long-term occupancy of competing attractors. We present a pedagogical, physics-oriented review built on Feng--Kurtz's unified Hamiltonian framework: from the infinitesimal generator of a Markov process we derive the Hamiltonian, the Lagrangian and action via Legendre--Fenchel duality, and the quasi-potential. The same formalism covers weak-noise diffusions, Markov jump processes, interacting particle systems, and nonequilibrium stochastic fields. We show how broken detailed balance separates forward and reverse transition paths, and how $W$-graphs construct multi-attractor landscapes. Finally, we connect the theory to numerical methods, including ordered upwind methods, geometric minimum action algorithms, and neural-network parameterizations, and provide an accompanying GPU-accelerated Python package.
\end{abstract}

\maketitle

\section{Introduction}
\label{sec:intro}

Many systems in statistical physics, biophysics, and nonlinear dynamics spend the overwhelming majority of their time dwelling near one of several long-lived metastable states, transitioning between them only on rare occasions. Activated open chemical reactions \cite{degn_bistability_1968, schlogl_chemical_1972, geiseler_three_1977, bachmann_autocatalytic_1992, semenov_autocatalytic_2016}, switching events in genetic and biochemical networks \cite{novick_enzyme_1957, monod_general_1961, ptashne_genetic_1986, gardner_construction_2000, craciun_understanding_2006, thomson_unlimited_2009}, nucleation phenomena \cite{hanggi_reaction-rate_1990}, protein conformational changes \cite{shaw_atomic-level_2010, shan_transitions_2013}, catastrophic regime shifts in ecosystems \cite{nolting_balls_2016}, traveling waves in neural circuits \cite{nartallo-kaluarachchi_nonequilibrium_2026, hancock_metastability_2025}, and transitions between competing spatial patterns or active phases \cite{cross_pattern_1993, fogedby_minimum_2009, zakine_unveiling_2024, cates_motility-induced_2015, cates_active_2025, zwicker_physics_2025} all share this common architecture. The rarity of these transitions by no means diminishes their physical importance. On sufficiently long time scales, it is precisely these rare events that select which state is predominantly observed, govern how hysteresis is eventually resolved, dictate how the system responds to perturbations, and determine the lifetime of metastable configurations.

Throughout this review, our primary object of study is a time-homogeneous Markov process $X^\eps_t\in\RR^{d}$ whose stochastic fluctuations are controlled by a single small parameter $\eps$, and the goal is to characterize the asymptotic behavior of the system in the small $\eps$ limit. A canonical representative is the It\^o stochastic differential equation (SDE):
\begin{equation}
\label{eq:generic_sde}
    dX_t^{\eps} = b(X_t^{\eps})\,dt + \sqrt{\eps}\,\sigma(X_t^{\eps})\,dW_t,
\end{equation}
where $b(\cdot)\in\RR^{d}$ denotes the deterministic drift vector field, $W_t$ is a standard Brownian motion in $\RR^{m}$, and $\sigma\in\RR^{d\times m}$ sets the amplitude and spatial geometry of the noise. While $\eps$ represents the noise strength in Eq.~\eqref{eq:generic_sde}, its physical origin varies across applications: in chemical master equations it corresponds to the inverse system volume $1/\Omega$; in systems of $N$ interacting particles it is $1/N$; and in thermal equilibrium systems it is $k_BT$. What unifies these diverse settings, and enables a universal theoretical treatment, is that in the limit $\eps\to0$ the dynamics become deterministic. On ordinary time scales, trajectories of Eq.~\eqref{eq:generic_sde} converge to solutions of the deterministic dynamical system
\begin{equation}
    \dot x = b(x).
    \label{eq:deterministic_dynamics}
\end{equation}

If Eq.~\eqref{eq:deterministic_dynamics} has several attractors, the noise can nevertheless carry the system from one basin to another. Such transitions become exponentially rare as $\eps\to0$, and the reason is worth seeing immediately, because the same calculation already contains the whole theory in miniature.

\paragraph*{Why transitions are exponentially rare as $\eps\to0$.}
Over a short time interval $\Delta t$, the increment of Eq.~\eqref{eq:generic_sde} is $\Delta X = b\,\Delta t + \sqrt\eps\,\sigma\,\Delta W$, where $\Delta W \sim \mathcal N(0,\,\Delta t\,\mathbb{I})$ is a standard Gaussian random vector. If the state moves with an instantaneous velocity $v$ differing from the deterministic drift $b$, the discrepancy $v-b$ must be supplied entirely by the noise:
\begin{equation}
    (v-b)\,\Delta t \approx \sqrt{\eps}\,\sigma\,\Delta W 
    \qquad\Longleftrightarrow\qquad
    \Delta W \approx \frac{u\,\Delta t}{\sqrt\eps},
    \qquad u \equiv \sigma^{-1}(v-b).
\end{equation}
The Gaussian probability density of realizing an increment of this magnitude scales as
\begin{equation}
    \propto \exp\!\left(-\frac{|\Delta W|^2}{2\Delta t}\right)
    = \exp\!\left(-\frac{|u|^2\,\Delta t}{2\eps}\right).
\end{equation}
Because increments across non-overlapping intervals are statistically independent, their probabilities multiply. Compounding these factors along a continuous trajectory over the time window $0\leqslant t\leqslant T$ yields the path probability:
\begin{equation}
    \label{eq:intro_heuristic_action}
    \PP\big[\text{path with velocity } \{\dot X_t^{\eps}\}_{0\leqslant t\leqslant T}\big]
    \;\asymp\;
    \exp\!\left(-\frac{1}{2\eps}\int_0^T \big|\sigma^{-1}(\dot X_t^{\eps} - b(X_t^{\eps}))\big|^2\,dt\right).
\end{equation}
Here the symbol ``$\asymp$'' denotes logarithmic equivalence in the limit $\eps\to0$:
\begin{equation}
    \label{eq:asymp_def}
    \PP_\eps(A) \;\asymp\; e^{-I_A/\eps}
    \qquad\text{means precisely}\qquad
    \lim_{\eps\to0} \eps \log \PP_\eps(A) \;=\; -I_A .
\end{equation}
Three fundamental insights follow at once. First, any path that leaves the basin of attraction must push against the deterministic flow over a duration of $O(1)$; consequently, the integral in Eq.~\eqref{eq:intro_heuristic_action} is strictly positive and the exponent scales as $O(1/\eps)$, rendering the transition exponentially rare. Second, the exponent is \emph{already an action functional}---a time integral of a non-negative Lagrangian cost density that vanishes identically if and only if $\dot X_t^{\eps} = b(X_t^{\eps})$. Third, the central physical question shifts from ``how probable is a transition?'' to ``which path minimizes this action integral?'', because in an exponential competition, the minimizing path dominates all competing fluctuations by an exponentially overwhelming margin. Everything in this review is an elaboration of Eq.~\eqref{eq:intro_heuristic_action}: how to systematically construct this action functional across diverse model classes, how to identify the path with the minimum action, and what physical principles govern global stability and rare transitions.

\paragraph*{What deterministic analysis cannot tell us.}
Linear stability analysis identifies fixed points, limit cycles, and basin boundaries, characterizing how rapidly a system recovers from an infinitesimal perturbation. It cannot answer any of the following:
\begin{enumerate}
    \item \textbf{Which attractor is globally ``more stable''?} That is, where does the stochastic system spend the majority of its time over asymptotically long observation windows?
    \item \textbf{Along which trajectory does a transition occur?} In the weak-noise limit $\eps\to0$, rare transitions concentrate tightly along a single \emph{optimal transition path} (or a small set of them).
    \item \textbf{How long is the mean escape time?} What determines the exponential scale governing the waiting time between transition events?
\end{enumerate}
In thermal equilibrium gradient systems, these questions have classical answers. If the drift satisfies $b=-\nabla U$ for some scalar potential energy $U(x)$, and the noise is isotropic with an Einstein fluctuation--dissipation relation, the potential landscape $U$ completely orders the relative stability of states: potential barriers dictate escape rates via the Arrhenius law, and an uphill fluctuation is the exact time reversal of deterministic relaxation. The transition pathway is then simply the minimum energy path passing through the lowest intervening saddle point \cite{hanggi_reaction-rate_1990, e_minimum_2004, amir_thinking_2020}.

Away from equilibrium, however, this classical picture breaks down completely. The drift field $b$ is generally non-conservative ($\nabla \times b \neq 0$); thus, persistent stationary probability currents circulate in state space, forward and reverse transition pathways no longer coincide, and the very notion of a saddle point's ``height'' becomes ill-defined because no global scalar potential exists. Furthermore, the noise covariance matrix $D=\sigma\sigma^\top$ can also play a decisive role: directions that are cheap to explore under isotropic thermal noise may be prohibitively expensive, or strictly inaccessible, under anisotropic, state-dependent, or degenerate noise.

\begin{figure}[tbp]
    \centering
    \includegraphics[width=\textwidth, keepaspectratio]{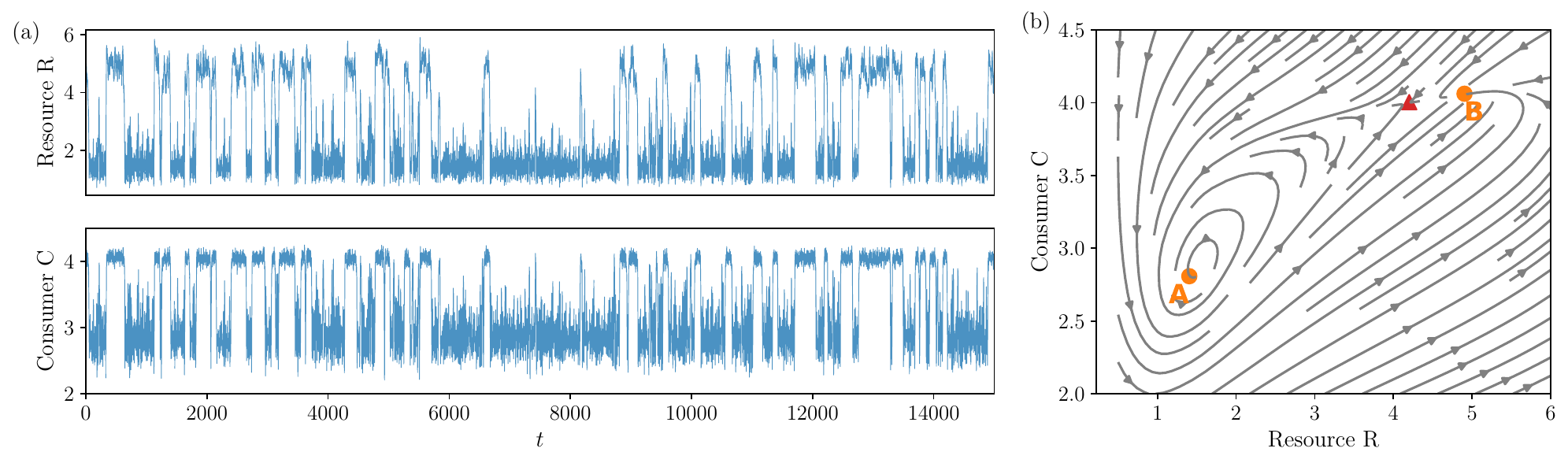}
    \caption{Local stability does not determine global stability in nonequilibrium systems. (a) Stochastic time series of resource $R_t$ and consumer $C_t$ displaying metastable dwelling punctuated by rare, noise-induced transitions between attractors $A$ and $B$. (b) Phase portrait in the $(R,C)$ plane showing deterministic streamlines (grey), stable attractors $A$ and $B$ (orange circles), and the intervening saddle point (red triangle). Although state $B$ exhibits faster local recovery following small perturbations, the system globally spends the majority of its time dwelling near $A$. The parameters used in the simulation are: $\alpha = 1.54$, $\beta = 10.14$, $\gamma = 0.476$, $\delta = 1$, $\mu = 0.112509$, $\kappa=1$, $\sigma_R=0.1$, and $\sigma_C = 0.05$. The two stable fixed points are $A\approx(1.405, 2.808)$ and $B\approx(4.904, 4.062)$.}
    \label{fig:consumer_resource}
\end{figure}

\paragraph*{Why linear stability analysis cannot determine global stability.}
In linear stability analysis, one linearizes the deterministic flow around a fixed point and evaluates local decay rates from the eigenvalues of the Jacobian matrix \cite{strogatz_nonlinear_2018}. Crucially, this reflects only local curvature and recovery. To see how dramatically local and global stability can diverge, consider a stochastic consumer--resource ecosystem exhibiting a Holling type-III functional response \cite{nolting_balls_2016}, standard in modeling phytoplankton--zooplankton population dynamics. Letting $R_t$ and $C_t$ denote the resource and consumer biomasses, their coupled dynamics are described by the SDEs:
\begin{align}
    \label{eq:consumer-resource}
    dR_t & = \left[ \alpha R_t\left(1-\frac{R_t}{\beta}\right) -\delta\,\frac{R_t^2 C_t}{\kappa+R_t^2} \right]dt +\sigma_R\,dW_t^{(R)}, \qquad
    dC_t = \left[ \gamma\,\frac{R_t^2 C_t}{\kappa+R_t^2} -\mu C_t^2 \right]dt +\sigma_C\,dW_t^{(C)}.
\end{align}
In the absence of noise, the system is bistable across a broad parameter regime (see Fig.~\ref{fig:consumer_resource}) with two coexisting stable fixed points, $A$ and $B$. Locally, state $B$ appears significantly more robust: its slowest recovery rate (determined by the spectral abscissa of the Jacobian, i.e., the smallest $|\mathrm{Re}\,\lambda|$) is $0.0927$, nearly double the value of $0.0474$ at $A$. Hence, an infinitesimal perturbation decays roughly twice as fast near $B$.\footnote{Comparing a pair of real eigenvalues with a complex conjugate pair requires care: the appropriate scalar comparison is between the spectral abscissas $\max_i\mathrm{Re}\,\lambda_i$, which govern the asymptotic local recovery rate, rather than between individual eigenvalue magnitudes.} Globally, however, the ranking completely inverts: an empirical tally from a long numerical trajectory demonstrates that the stochastic system spends over $64\%$ of its time dwelling near attractor $A$ and only $\approx 36\%$ near $B$. Fig.~\ref{fig:consumer_resource} illustrates both the stochastic time series and the corresponding phase plane.

Strictly speaking, the simulation parameters above lie outside the asymptotic weak-noise regime: the noise amplitudes $\sigma_R=0.1$ and $\sigma_C = 0.05$ were chosen sufficiently large to observe multiple transitions within an accessible computational window. Yet this limitation highlights the very purpose of Freidlin--Wentzell theory: when noise is genuinely weak, transitions become astronomically rare and brute-force simulation becomes impossible. One thus requires a theoretical and computational framework that yields transition rates, pathways, and global weights directly from the action exponent without ever having to wait for a rare event to occur spontaneously.

\paragraph*{What Freidlin--Wentzell theory delivers.}
The theory developed by Freidlin and Wentzell \cite{wentzell_small_1969, wentzell_limiting_1969, wentzell_small_1970, freidlin_random_2012} resolves all three questions posed above at the level of logarithmic asymptotics. To every continuous trajectory $\phi$, it assigns an action functional $S_T[\phi]$ such that the probability of observing paths near $\phi$ obeys $\PP[\phi]\asymp \exp\left(-S_T[\phi]/\eps\right)$. Minimizing this action over all paths connecting two given states defines the \emph{quasi-potential} $V(x_a, x_b)$, which generalizes the equilibrium potential energy $U$ to arbitrary non-gradient systems. The optimal path realizing this minimum is the \emph{minimum action path}, also known as the \emph{instanton}; the quasi-potential barrier height dictates the exponential scale of the mean escape time; and a combinatorial construction over the attractors (optimal $W$-graphs) converts pairwise barriers into global stationary weights in order to construct a global quasi-potential $V(x)$ such that the stationary distribution $\PP_\eps(x) \;\asymp\; \exp\left(-V(x)/\eps\right)$. In this way, the relative occupancy ratio of competing attractors is determined rigorously from large deviation principles.

\subsection{Scope of the review}
\label{sec:scope}

A rich mathematical literature treats Freidlin--Wentzell theory and general large deviation principles \cite{varadhan_large_2008, ellis_entropy_2006, dembo_large_2010, freidlin_random_2012, feng_large_2006, touchette_introduction_2018, touchette_large_2009}. However, existing treatments are predominantly measure-theoretic treatises aimed at mathematicians, or focus primarily on the long-time dynamical large deviations of time-averaged observables, i.e., the large deviation properties of quantities like $\frac{1}{T}\int_0^T f(x_t)dt$ (often referred to as the Donsker--Varadhan regime). This review is written specifically for physicists and applied scientists who seek to \emph{apply} Freidlin--Wentzell theory to concrete models without navigating measure-theoretic technicalities.

Our presentation is deliberately heuristic and physical. We exchange limits when the physics is transparent, speak of probability densities where a mathematician would specify Borel sets, use the term ``path'' to represent a narrow tube in trajectory space, and focus primarily on the dominant exponential action. While we do not delve deeply into subexponential prefactors, we discuss in Sec.~\ref{sec:fw_limits} the physical regimes where leading-order theory requires correction and point to modern methods for evaluating fluctuation determinants \cite{hanggi_reaction-rate_1990, bouchet_path_2022, heller_evaluation_2024, schorlepp_systematic_2025}. In addition, though Freidlin--Wentzell theory applies to discrete-time Markov processes \cite{kifer_discrete-time_1990}, in this review we focus on continuous-time cases only.

The organizing thread is a simplified version of the modern Hamiltonian viewpoint, systematized by Feng and Kurtz \cite{feng_large_2006}:
\begin{equation*}
    \text{Infinitesimal generator }\mathcal L
    \;\rightarrow\;
    \text{Hamiltonian } H(x,k)
    \;\rightarrow\;
    \text{Lagrangian }L(x,\dot x)
    \;\rightarrow\;
    \text{action functional }S_T
    \;\rightarrow\;
    \text{quasi-potential } V .
\end{equation*}

On the computational side, we complement the theoretical exposition with a high-performance Python package featuring GPU acceleration, implementing standard and gradient-based (automatic-differentiation) algorithms for computing optimal transition paths, quasi-potential landscapes, and $W$-graph weights.

\subsection{A short lineage}
\label{sec:lineage}

Several distinct scientific traditions converged upon the action, Hamiltonian, WKB, and quasi-potential formalism from different perspectives, leading to terminology that is often redundant across disciplines. A brief orientation helps place these concepts in their historical context.

On the physics side, Kramers' foundational 1940 work established that escape across an energy barrier is governed by an exponential Arrhenius factor modulated by a dynamical prefactor \cite{kramers_brownian_1940}, launching decades of reaction-rate theory reviewed in \cite{hanggi_reaction-rate_1990}. Onsager and Machlup subsequently formulated a path-probability functional for linear Gaussian processes \cite{onsager_fluctuations_1953}, representing the linear, Gaussian special case of the action in Eq.~\eqref{eq:intro_heuristic_action}. Kubo and collaborators extended these ideas to macroscopic variables governed by master equations \cite{kubo_fluctuation_1973}.

On the mathematics side, Varadhan's pioneering work on small-time diffusion asymptotics and exponential integrals \cite{varadhan_diffusion_1967, varadhan_large_2008} provided the rigorous foundation of large deviation theory. Building on this foundation, Freidlin and Wentzell developed a systematic theory of small random perturbations of dynamical systems: establishing path-space large deviation principles, formulating exit problems and exit boundary distributions, defining the quasi-potential, and creating the graph-theoretic constructions for multi-attractor metastability that we present in Sec.~\ref{sec:multiple_attractors} \cite{wentzell_small_1969, wentzell_limiting_1969, wentzell_small_1970, freidlin_random_2012}\footnote{Wentzell's name is also transliterated as Ventcel' or Ventsel' in older translations.}.

In physics, the reception occurred in two subsequent waves. Ludwig's influential 1975 review introduced WKB methods for extinction and persistence to applied scientists and ecologists \cite{ludwig_persistence_1975}. Graham and T\'el systematically constructed ``nonequilibrium potentials'' for dissipative dynamical systems, discovering that such potentials are generically non-differentiable due to the emergence of caustics and switching lines where distinct optimal paths collide \cite{graham_existence_1984, graham_weak-noise_1985, graham_nonequilibrium_1991}. In parallel, eikonal and WKB approximations for master equations became standard in chemical kinetics and population biology \cite{dykman_large_1994, elgart_rare_2004, assaf_wkb_2017}, and closely analogous structures underpin the macroscopic fluctuation theory of driven diffusive systems \cite{bertini_macroscopic_2015, falasco_macroscopic_2025}.

Finally, Feng and Kurtz showed that these disparate developments are specific manifestations of a single mathematical structure: the convergence of logarithmically transformed nonlinear semigroups, governed by a limiting nonlinear generator $Hf(x)=H(x,\nabla f(x))$ whose associated Hamilton--Jacobi equation is understood in the viscosity sense\footnote{In brief, first-order Hamilton--Jacobi equations generically fail to possess smooth solutions everywhere because characteristics (optimal paths) intersect, generating ridges, kinks, or caustics where spatial derivatives cease to exist. A \emph{viscosity solution} is the unique, continuous weak solution selected by the zero-noise limit (analogous to vanishing viscosity in fluid dynamics). Whenever multiple transition pathways compete, the viscosity solution automatically selects the branch with the globally minimal action.} \cite{feng_large_2006}. Feng and Kurtz unified Hamiltonian large deviations across virtually arbitrary model classes. That fundamental unification is what makes possible the universal approach developed throughout this review.

\subsection{Notation and Outline}
\label{sec:notation}

Two conventions warrant special emphasis because notations in the literature vary widely. First, we denote the conjugate momentum by $k$ (whereas many mathematical texts employ $p$ or $\theta$). Second, we consistently define the diffusion matrix as $D=\sigma\sigma^\top$ \emph{without} a factor of two, in accord with modern statistical physics literature \cite{touchette_introduction_2018, zakine_minimum-action_2023}; consequently, the infinitesimal generator associated with Eq.~\eqref{eq:generic_sde} carries a prefactor of $\eps D/2$. Angle brackets $\<\cdot,\cdot\>$ denote either the standard Euclidean inner product on $\RR^d$ or the $L^2$ inner product $\<k,\psi\>=\int k(r)\psi(r)\,dr$ when $k(r)$ and $\psi(r)$ are scalar fields. Finally, $\otimes$ denotes the outer product, $\nabla\nabla$ denotes the spatial Hessian operator and $A:B=\sum_{ij}A_{ij}B_{ij}$ represents the Frobenius inner product, so that $D:\nabla\nabla = \sum_{ij}D_{ij}\partial_i\partial_j$.


The remainder of this review is organized as follows. Section~\ref{sec:toolkit} establishes the core vocabulary and foundational theorems of large deviation theory. Section~\ref{sec:hamiltonian} defines the Hamiltonian as a local generating function, derives it directly from the infinitesimal generator, and establishes Legendre--Fenchel duality with the Freidlin--Wentzell Lagrangian. Section~\ref{sec:dictionary} constructs an explicit dictionary of Hamiltonians across major physical model classes. Section~\ref{sec:quasipotential} formulates the quasi-potential landscape, instanton paths, exit problems, and the combinatorial $W$-graph construction for multi-attractor metastability. Section~\ref{sec:numerics} reviews grid-based (ordered upwind) and path-based (geometric minimum action) numerical methods alongside their modern neural network extensions. Section~\ref{sec:discussion} concludes with open challenges and future outlook.

\section{The minimum toolkit of large deviations}
\label{sec:toolkit}

Freidlin--Wentzell theory is fundamentally a branch of Large Deviation Theory (LDT). Specifically, it is the theory of large deviations applied to path spaces of SDEs subject to small noise perturbations, or other types of Markov stochastic processes where a single parameter can control the stochasticity of the system, along with the dynamical consequences that follow from it. Large Deviation Theory is the overarching mathematical framework that quantifies the exponential decay of probabilities of rare events. In Varadhan's general formulation \cite{varadhan_diffusion_1967}, a family of probability measures $P_\varepsilon$ satisfies a Large Deviation Principle (LDP) with a non-negative function $I: \mathcal{X} \to [0, \infty]$ if:
$$P_\varepsilon(x) \asymp e^{- I(x)/\varepsilon}$$
Freidlin--Wentzell theory specializes this concept to continuous-time dynamical systems. The function $I(x)$ is often called the rate function in LDT literature, and the action functional in Freidlin--Wentzell theory.

\subsection{The mathematical minimum toolkit}
\label{sec:three_tools}

While the existence and uniqueness of the action functional can be established with rigorous measure theory, in this review we take a constructive and heuristic perspective. Remarkably, three elementary mathematical tools perform virtually all the analytical work throughout this review (for rigorous proofs, see \cite{dembo_large_2010, freidlin_random_2012}):

\paragraph*{Laplace's principle.} An exponential integral is asymptotically dominated in the limit $\eps\to0$ by the maximum of its exponent:
\begin{equation}
    \label{eq:laplace}
    \eps \log \int e^{f(x)/\eps} \, dx \;\longrightarrow\; \sup_x f(x)
    \qquad \mathrm{as}\; \eps\to0.
\end{equation}

\paragraph*{The contraction principle.} If $X^\eps$ satisfies a large deviation principle with rate function $I$, and $Y=f(X)$ is a continuous function of $X^\eps$, then $Y^\eps$ satisfies one as well, with
\begin{equation}
    \label{eq:contraction}
    I_Y(y) = \inf_{x:\,f(x)=y} I_X(x).
\end{equation}
In other words, the most likely way to realize a given macroscopic value $y$ is the least improbable among all microscopic configurations compatible with it. Every dimensional reduction performed in this review---from continuous paths to endpoints, from fixed-time transitions to the quasi-potential, and from microscopic coordinates to coarse-grained order parameters---is a direct application of Eq.~\eqref{eq:contraction}.

\paragraph*{Legendre--Fenchel duality.} For a convex function $L(x, v)$,
\begin{equation}
    \label{eq:legendre}
    H(x, k) = \sup_v \left\{\<k,v\> - L(x, v)\right\},
    \qquad
    L(x, v) = \sup_k \{\<k,v\> - H(x, k)\} .
\end{equation}
This is the familiar duality connecting the Lagrangian and Hamiltonian formulations of classical mechanics. While Freidlin and Wentzell originally formulated their results directly in terms of $L$, the Hamiltonian $H$ can be derived algebraically from the infinitesimal generator of the stochastic process, whereas $L$ often lacks an explicit closed form (e.g., in jump processes). We can therefore determine $H$ first and obtain $L$ via this duality.

\subsection{A path large deviation principle}
\label{sec:path_ldp}

Equipped with this vocabulary, the central assertion of Freidlin--Wentzell theory can be stated concisely. Let $X^\eps_{[0,T]}$ denote the stochastic trajectory over the time window $[0,T]$. Trajectories obey an LDP on path space with speed $1/\eps$ and action functional $S_T[\phi]$:
\begin{equation}
    \label{eq:path_ldp}
    \PP\big[X^\eps_{[0,T]} \approx \phi \big] \;\asymp\; \exp\left(-\frac{1}{\eps}S_T[\phi]\right),
    \qquad
    S_T[\phi] = \int_0^T L\big(\phi_t,\dot\phi_t\big)\,dt .
\end{equation}
Here the symbol $\approx$ signifies that the random trajectory remains confined within an arbitrarily narrow tube of radius $\delta$ surrounding $\phi$ under the uniform norm, with $S_T[\phi]=\infty$ for paths that are not absolutely continuous. For brevity, in the rest of the review we denote this probability simply as $\PP[\phi]$.

To evaluate the transition probability between two prescribed boundary states $x_a$ and $x_b$ in duration $T$, we apply the contraction principle, Eq.~\eqref{eq:contraction}, by minimizing over all paths satisfying $X^\eps_0 = x_a$ and $X^\eps_T = x_b$:
\begin{equation}
    \label{eq:ldp_path}
    \PP\big[X^\eps_T = x_b \;\big|\; X^\eps_0 = x_a\big] \;\asymp\; \exp\left(-\frac{1}{\eps}V_T(x_a,x_b)\right),
    \qquad
    V_T(x_a,x_b) = \inf_{\substack{\phi(0)=x_a\\ \phi(T)=x_b}} S_T[\phi].
\end{equation}
Eq.~\eqref{eq:ldp_path} defines the fixed-time optimal transition path, serving as the foundational variational objective for the numerical algorithms discussed in Sec.~\ref{sec:numerics}.

\emph{Remark.} Eq.~\eqref{eq:path_ldp} should not be confused with the Onsager--Machlup path functional \cite{onsager_fluctuations_1953}. The Onsager--Machlup theory includes subleading $O(1)$ terms, such as the divergence of the drift $\nabla\cdot b$, to describe the most probable tube of paths at finite noise, but is restricted to Gaussian diffusions and does not capture the asymptotic large deviation scaling for non-Gaussian or jump processes.

\section{The Feng--Kurtz formalism: from the generator to the Hamiltonian}
\label{sec:hamiltonian}

In this section, we formulate the Freidlin--Wentzell Lagrangian $L(x,\dot x)$ and Hamiltonian $H(x,k)$. We demonstrate that $H(x,k)$ can be evaluated directly from the infinitesimal generator of the Markov process and connected to $L(x,\dot x)$ via the Legendre--Fenchel transform.

\subsection{The cost of a velocity}
\label{sec:velocity_cost}

Following the heuristic calculation introduced in Sec.~\ref{sec:intro}, the displacement of a Markov process over a short time increment $\Delta t$ is approximately given by $\dot x\,\Delta t$, with an associated local probability scaling as
\begin{equation}
    \label{eq:velocity_ldp}
    \PP\big[\dot x_t \approx v \;\big|\; x_t = x\big] \;\asymp\; \exp\left(-\frac{1}{\eps} L(x,v)\,\Delta t\right) .
\end{equation}
For diffusions, this scaling was established explicitly in Eq.~\eqref{eq:intro_heuristic_action}; for general continuous-time Markov processes, it reflects the existence of an underlying local large deviation principle. This relation defines the \emph{Lagrangian} $L(x,v)$: the action cost per unit time of forcing the system to move with velocity $v$ while situated at position $x$. Because the deterministic velocity $b(x)$ is the most probable behavior (occurring with probability approaching one), all atypical velocities carry a cost: $L(x,v) \geqslant 0$, with equality $L(x,v)=0$ holding if and only if $v=b(x)$.

Because the increments of a Markov process across disjoint time intervals are conditionally independent, the probability of observing a given continuous path is the product of the probabilities of successive infinitesimal increments:
\begin{equation}
    \prod_{n=0}^{T/\Delta t}\PP\big[\dot x_{n\Delta t} \approx v_n \;\big|\; x_{n\Delta t}=x_n\big] \;\asymp\; \exp\left(-\frac{1}{\eps}\sum_n L(x_n,v_n)\,\Delta t\right)
\end{equation}

Taking the continuous-time limit $\Delta t\to0$, the sum in the exponent converges to the integral in Eq.~\eqref{eq:path_ldp}. We identify $S_T[\phi]=\int_0^T L(\phi_t,\dot\phi_t)\,dt$ as the \emph{action functional}. This is the central dynamical object of Freidlin--Wentzell theory.

This simple construction provides the conceptual foundation of path large deviations. What is missing, however, is a general operational method to compute $L(x,v)$. Direct evaluation via Gaussian increments works for simple diffusions, but fails for jump processes or interacting systems where $L$ cannot be integrated analytically. The standard recourse in statistical mechanics is to pass to the conjugate representation by evaluating the local cumulant generating function.

\subsection{The Hamiltonian as a local generating function}
\label{sec:H_as_scgf}

The Hamiltonian is defined as the scaled cumulant generating function of the local increment \cite{feng_large_2006}:
\begin{equation}
    \label{eq:hamiltonian_def}
    H(x, k) \;=\; \lim_{\eps \to 0}\;\lim_{\Delta t \to 0} \frac{\eps}{\Delta t}\,
    \log \EE\left[ \exp\left(\frac{1}{\eps}\<k,\, x_{t+\Delta t} - x_t\>\right) \;\Big|\; x_t = x\right] .
\end{equation}
The variable $k$ conjugate to the displacement represents the canonical momentum. Notice that Eq.~\eqref{eq:hamiltonian_def} is strictly ``local'', as the exponential tilt is applied to an infinitesimal increment at a fixed position $x$, in contrast to the time-extensive observables tilted in the Donsker--Varadhan long-time framework \cite{touchette_introduction_2018}.

To establish that $H$ is the Legendre transform of $L$, we substitute the velocity probability Eq.~\eqref{eq:velocity_ldp} into the expectation:
\begin{align}
    \EE\left[ e^{\<k,\dot{x}\>\Delta t/\eps} \right]
    & \asymp \int e^{\<k,\dot{x}\>\Delta t/\eps}\, e^{-L(x, \dot{x}) \Delta t/\eps} \, d\dot{x}
    = \int e^{\{\<k,\dot{x}\> - L(x, \dot{x})\} \Delta t/\eps} \, d\dot{x} ,
\end{align}
and apply Laplace's principle, Eq.~\eqref{eq:laplace}, with $\Delta t/\eps$ serving as the large parameter:
\begin{equation}
    \frac{\eps}{\Delta t}\log\EE\left[ e^{\<k,\dot{x}\>\Delta t/\eps} \right]
    \;\longrightarrow\; \sup_{v}\left\{\<k,v\> - L(x,v)\right\}.
\end{equation}
Comparing with Eq.~\eqref{eq:hamiltonian_def},
\begin{equation}
    \label{eq:hamiltonian_legendre}
    H(x,k) = \sup_{v}\big\{\<k,v\> - L(x,v)\big\},
    \qquad
    L(x,v) = \sup_{k}\big\{\<k,v\> - H(x,k)\big\} .
\end{equation}
The second equality represents the inverse transform, guaranteed by the convexity of $L$ in $v$ (Sec.~\ref{sec:convexity}).

Two general structural properties follow from Eq.~\eqref{eq:hamiltonian_def} and serve as universal consistency checks for any model class:
\begin{enumerate}
    \item $H(x,0)=0$, reflecting the exact conservation and normalization of probability at zero tilt.
    \item The gradient with respect to momentum at zero tilt yields the deterministic velocity:
\begin{equation}
    \label{eq:drift_from_H}
    \left.\frac{\partial H}{\partial k}\right|_{k=0} = b(x),
\end{equation}
    because the first derivative of a cumulant generating function generates the mean.
\end{enumerate}
The portion of $H(x,k)$ linear in $k$ always encodes the deterministic dynamics, while the non-linear terms in $k$ quantify the cost of stochastic fluctuations. As we show below, this clean separation holds universally across diffusions, jump processes, and field theories.

Eq.~\eqref{eq:hamiltonian_def} can also be expressed through the infinitesimal generator $\mathcal{L}_\eps$ of the Markov process whose stochasticity is controlled by $\eps$\footnote{We write $\mathcal{L}_\eps$ instead of $\mathcal{L}$ to emphasize that $\mathcal{L}$ depends on $\eps$.}. Recall that $\mathcal L_\eps$ is defined by its action on smooth test functions $f$:
\begin{equation}
    \lim_{t \downarrow 0} \frac{\EE[f(X_t)\,|\,X_0 = x] - f(x)}{t} = (\mathcal{L}_\eps f)(x),
    \qquad\text{so that}\qquad
    \EE[f(X_{t+\Delta t})\,|\,X_t = x] \approx f(x) + \Delta t\,(\mathcal{L}_\eps f)(x).
\end{equation}
Choose the exponential test function $f(x) = e^{\<k, x\>/\eps}$. Then
\begin{equation}
    \EE\left[e^{\<k, X_{t+\Delta t}\>/\eps} \,\big|\, X_t = x\right]
    \approx e^{\<k, x\>/\eps} + \Delta t\, \big(\mathcal{L}_\eps\, e^{\<k, \cdot\>/\eps}\big)(x),
\end{equation}
and dividing by $e^{\<k,x\>/\eps}$, taking the logarithm and using $\log(1+\delta)\approx\delta$,
\begin{equation}
    \frac{\eps}{\Delta t}\,\log \EE\left[e^{\<k,\,X_{t+\Delta t} - x\>/\eps} \,\big|\, X_t = x\right]
    \approx \eps\, e^{-\<k, x\>/\eps}\, \big(\mathcal{L}_\eps\, e^{\<k, \cdot\>/\eps}\big)(x).
\end{equation}
Hence
\begin{equation}
    \label{eq:hamiltonian_generator}
    \boxed{\;
    H(x,k) = \lim_{\eps \to 0}\; \eps\, e^{-\<k, x\>/\eps}\;
    \mathcal{L}_\eps\left( e^{\<k,\,\cdot\>/\eps} \right)(x) \; }
\end{equation}

\emph{Remark on the order of limits.} In Eq.~\eqref{eq:hamiltonian_def}, the limit $\Delta t\to0$ is taken prior to $\eps\to0$, whereas the heuristic derivation of Sec.~\ref{sec:velocity_cost} considers $\Delta t$ small but fixed as $\eps\to0$. For all physical models treated in this review, the two limits commute. Subtle discrepancies can arise only in systems with widely separated internal relaxation time scales or where transition rates vanish on boundaries of state space.

\subsection{Back to the action}
\label{sec:back_to_action}

With the Hamiltonian $H(x,k)$ in hand, the Lagrangian follows immediately from the Legendre transform, Eq.~\eqref{eq:hamiltonian_legendre}, and the action is obtained by time integration: $S_T[\phi]=\int_0^T L(\phi,\dot\phi)\,dt$. The variational problem $\delta S_T = 0$ yields the Euler--Lagrange equations of motion, or equivalently, Hamilton's canonical equations:
\begin{equation}
    \label{eq:hamilton_equations}
    \dot x = \frac{\partial H}{\partial k}, \qquad \dot k = -\frac{\partial H}{\partial x} .
\end{equation}

\subsection{Remarks}
\label{sec:convexity}

\emph{Convexity}. For each fixed $x$, $H(x,k)$ is a limit of scaled logarithms of moment generating functions, and such functions are convex in $k$ by H\"older's inequality. Since pointwise limits of convex functions are convex, $H$ is always convex in $k$, and $L=H^\ast$ is always convex in $v$, with $H=L^\ast$ by biconjugacy. This is why the Legendre--Fenchel transform Eq.~\eqref{eq:hamiltonian_legendre} is valid in both directions. Note that $H$ does not need to be \emph{strictly} convex, in which case $L$ has flat directions or is infinite outside a cone.

\emph{Degenerate noise.} When $\sigma$ is not square or not invertible, $\sigma^{-1}$ in Eq.~\eqref{eq:intro_heuristic_action} is meaningless and some velocities cost infinitely much; see Eq.~\eqref{eq:degenerate_lagrangian}. The Hamiltonian remains perfectly well defined, which is another practical argument for working on the Hamiltonian side.

\emph{Order of limits.} Weak noise, large system, continuum, and long-time limits need not commute, and a calculation that takes them in a different order from the experiment it models may describe a different physical regime.

\section{Hamiltonians for Canonical Model Classes}
\label{sec:dictionary}

We now apply the generator formula, Eq.~\eqref{eq:hamiltonian_generator}, to derive explicit Hamiltonians across several foundational model classes that commonly appear in nonequilibrium statistical physics, biophysics, and ecology \cite{zakine_minimum-action_2023}.

\subsection{Weak-noise diffusions}
\label{sec:diffusion}

Consider a multi-dimensional diffusion process described by the It\^o SDE:
\begin{equation}
    \label{eq:sde}
    dX_t = b(X_t)\,dt + \sqrt{\eps}\,\sigma(X_t)\,dW_t ,
\end{equation}
where $b:\RR^d\to\RR^d$ is the drift, $\sigma:\RR^d\to\RR^{d\times m}$ is the noise matrix, and $W_t$ is an $m$-dimensional standard Brownian motion. By It\^o's lemma (Appendix~\ref{appendix:generator}), the infinitesimal generator is
\begin{equation}
    \label{eq:sde_generator}
    \mathcal{L}_\eps = b(x)\cdot\nabla + \frac{\eps}{2}\,D(x):\nabla\nabla,
    \qquad D(x)=\sigma(x)\sigma(x)^\top .
\end{equation}
Inserting this into Eq.~\eqref{eq:hamiltonian_generator}, and noting that $\nabla e^{\<k,x\>/\eps} = (k/\eps)e^{\<k,x\>/\eps}$ and $\nabla\nabla e^{\<k,x\>/\eps} = (k\otimes k/\eps^2)e^{\<k,x\>/\eps}$,
\begin{align}
    H(x,k) & = \lim_{\eps \to 0}
    \left\{\eps\, e^{-\<k, x\>/\eps} \, b(x)\cdot \nabla e^{\<k, x\>/\eps}
    + \eps\,e^{-\<k, x\>/\eps}\, \frac{\eps}{2} \big[D(x):\nabla\nabla e^{\<k, x\>/\eps}\big]\right\} \nonumber \\
    & = \<k, b(x)\> + \frac{1}{2} D(x):(k\otimes k) ,
\end{align}
that is,
\begin{equation}
    \label{eq:hamiltonian_diffusive}
    \boxed{\;H(x,k) = \<k, b(x)\> + \tfrac{1}{2} \<k, D(x)\, k\>\;}
\end{equation}
Both terms are exactly as predicted in Sec.~\ref{sec:H_as_scgf}: linear in $k$ is the deterministic drift, quadratic in $k$ is the fluctuation cost.

The corresponding Lagrangian is obtained via the Legendre transform, Eq.~\eqref{eq:hamiltonian_legendre}. The supremum over $k$ is attained at the conjugate momentum $k^*$ satisfying
\begin{align}
    \partial_k\left\{\< k,\dot{x} \> - H(x,k)\right\} = 0 \longrightarrow k^* = [\sigma(x)\sigma(x)^\top]^{-1}(\dot{x} - b(x)),
\end{align}
therefore
\begin{align}
    \label{eq:lagrangian_diffusive}
    L(x, \dot{x}) = \frac{1}{2}\left|\sigma(x)^{-1}\left(\dot{x} - b(x)\right)\right|^2.
\end{align}
This is the classical Freidlin--Wentzell Lagrangian \cite{freidlin_random_2012}, and it reproduces the heuristic Eq.~\eqref{eq:intro_heuristic_action} in the introduction.

\paragraph*{Degenerate or non-square noise.} The second form in Eq.~\eqref{eq:lagrangian_diffusive} presupposes that $\sigma$ is square and invertible. In general it is not, e.g., in an inertial system where noise acts only on the momenta, or in a reaction network with fewer independent noise sources than species. The correct statement is a control representation: the cost is the cheapest noise that produces the required velocity,
\begin{equation}
    \label{eq:degenerate_lagrangian}
    L(x,v) = \inf\left\{ \tfrac12 |u|^2 \;:\; v = b(x) + \sigma(x)\,u \right\}
    = \begin{cases}
        \tfrac12\,\<v-b,\; D^{+}(v-b)\>, & v - b \in \ran \sigma(x), \\[2pt]
        +\infty, & \text{otherwise},
    \end{cases}
\end{equation}
with $D^{+}$ the Moore--Penrose pseudoinverse. Velocities outside the range of $\sigma$ are infinitely expensive \emph{instantaneously}, but this does not mean they are unreachable: the drift can transport the system through directions the noise cannot excite, so finite-action paths through such regions still exist, built from segments that combine affordable kicks with free drift. Note that $H$ in Eq.~\eqref{eq:hamiltonian_diffusive} needs no modification whatsoever; it is merely no longer strictly convex.

\subsection{Markov jump processes and the chemical master equation}
\label{sec:jump}

Consider a Markov jump process on integer states, as for a chemical reaction network,
\begin{equation}
    \label{eq:jump_process}
    \frac{\partial P(X, t)}{\partial t} = \sum_{j=1}^{R} \big[ W_j(X - \nu_j)\, P(X - \nu_j, t) - W_j(X)\, P(X, t) \big],
\end{equation}
where $R$ is the number of reaction channels, $\nu_j\in\ZZ^d$ the stoichiometric change vector of channel $j$, and $W_j(X)\geqslant0$ its propensity (reaction rate). Introduce the system size $\Omega$, e.g., the volume or the total particle number, and the intensive variable $x=X/\Omega$, and assume the standard scaling
\begin{equation}
    W_j(X) = \Omega\, w_j(x) + O(1) .
\end{equation}
A single reaction changes $x$ by $\nu_j/\Omega$, so jumps become small while their rate becomes large---the structure that makes a large-deviation limit possible. The small parameter is $\eps=1/\Omega$, and the generator acting on a function of the intensive variable is
\begin{equation}
    \label{eq:jump_generator}
    \mathcal{L}_\Omega f(x) = \sum_{j=1}^R \Omega\, w_j(x) \big[ f(x + \nu_j/\Omega) - f(x) \big] .
\end{equation}
Eq.~\eqref{eq:jump_generator} is the Kolmogorov backward equation \cite{gardiner_stochastic_2009}. Applying Eq.~\eqref{eq:hamiltonian_generator} with $\eps=1/\Omega$ and test function $e^{\Omega\<k,x\>}$,
\begin{align}
    H(x,k) & = \lim_{\Omega \to \infty} \frac{1}{\Omega} \, e^{- \Omega\<k, x\>} \sum_{j=1}^R \Omega\, w_j(x)
    \left[ e^{\Omega\<k,\, x+ \nu_j/\Omega\>} - e^{\Omega\<k, x\>} \right] \nonumber \\
    & = \lim_{\Omega \to \infty} \frac{1}{\Omega} \sum_{j=1}^R \Omega\, w_j(x)\left[ e^{\<k, \nu_j\>} - 1 \right],
\end{align}
so all $\Omega$ are cancelled, and we have:
\begin{equation}
    \label{eq:hamiltonian_jump_process}
    \boxed{\;H(x,k) = \sum_{j=1}^R w_j(x) \left( e^{\<k,\, \nu_j\>} - 1 \right)\;}
\end{equation}

\paragraph*{Relation to van Kampen's $\Omega$-expansion.} The derivation resembles van Kampen's system-size expansion \cite{van_kampen_power_1961}, but the two describe different parts of the distribution. The $\Omega$-expansion posits $X_t = \Omega x_t + \Omega^{1/2}\xi_t$ with $x_t$ the deterministic solution and $\xi_t$ a Gaussian fluctuation. This is a central-limit statement and produces the linear noise approximation. Eq.~\eqref{eq:hamiltonian_jump_process} posits $P\asymp e^{-\Omega I}$ and makes no Gaussian assumption. Expanding Eq.~\eqref{eq:hamiltonian_jump_process} to second order in $k$, which is the regime of small deviations, recovers the diffusive Hamiltonian:
\begin{equation}
    H(x,k) \approx \<k,\, b(x)\> + \tfrac{1}{2}\<k, D(x)\,k\>,
    \qquad b(x)=\sum_j w_j(x)\,\nu_j, \quad D(x) = \sum_j w_j(x)\,\nu_j\nu_j^\top .
\end{equation}
So the linear noise approximation is the small-$k$, i.e., small-deviation, corner of the exact theory.

\paragraph*{Why the difference is exponentially large: an exactly solvable case.} Take a one-dimensional birth--death process with rates $w_\pm(x)$ and jumps $\nu_\pm=\pm1$, so that
\begin{equation}
    H(x,k) = w_+(x)\big(e^{k}-1\big) + w_-(x)\big(e^{-k}-1\big).
\end{equation}
The Legendre transform can be done in closed form. Stationarity gives $v = w_+e^k - w_-e^{-k}$, therefore $e^{k^\ast} = \sqrt{w_-/w_+}\;e^{\theta}$ with $\theta = \operatorname{arcsinh}\!\big(v/2\sqrt{w_+w_-}\big)$, and
\begin{equation}
    \label{eq:birth_death_lagrangian}
    L(x,v) = v\left[\operatorname{arcsinh}\frac{v}{2\sqrt{w_+w_-}} + \frac{1}{2}\log\frac{w_-}{w_+}\right]
    - \sqrt{v^2+4w_+w_-} + w_+ + w_- .
\end{equation}
Clearly, this is not quadratic in $v$; for large $|v|$ it grows as $|v|\log|v|$ rather than $v^2$. This is because moving at speed $v$ requires of order $\Omega v$ net reaction events per unit time, and the tail of a Poisson variable decays like $e^{-n\log n}$ rather than $e^{-n^2}$, so the Gaussian approximation charges too much for fast excursions and systematically underestimates their probability.

\subsection{Stochastic fields}
\label{sec:fields}

Consider a scalar field $\psi(r,t)$ obeying the following stochastic partial differential equation (SPDE):
\begin{equation}
    \label{eq:spde}
    \frac{\partial\psi(r,t)}{\partial t} = A[\psi](r,t) + \sqrt{\eps}\,\xi(r,t),
\end{equation}
with $A[\psi]$ a deterministic drift functional and $\xi$ a Gaussian noise with field-dependent spatial correlation
\begin{equation}
    \EE\big[\xi(r,t)\,\xi(r',t')\big] = Q[\psi](r,r')\,\delta(t-t') .
\end{equation}
For a smooth functional $G[\psi]$, It\^o's lemma reads
\begin{equation}
    dG = \int\! dr\,\frac{\delta G}{\delta \psi(r)}\,d\psi(r)
    + \frac{1}{2} \int\! dr \int\! dr'\, \frac{\delta^2 G}{\delta \psi(r)\, \delta \psi(r')}\, d\psi(r)\, d\psi(r'),
\end{equation}
and with $d\psi(r)\,d\psi(r') = \eps\, Q(r,r')\,dt$, the infinitesimal generator is
\begin{equation}
    \label{eq:spde_generator}
    \mathcal{L}_\eps = \int\! dr\, A[\psi](r)\, \frac{\delta}{\delta \psi(r)}
    + \frac{\eps}{2}\int\! dr \int\! dr'\, Q(r,r')\, \frac{\delta^2}{\delta \psi(r)\, \delta \psi(r')} .
\end{equation}
The Hamiltonian follows from Eq.~\eqref{eq:hamiltonian_generator} exactly as in the finite-dimensional case. The test function becomes the test functional $G[\psi]=e^{\<k,\psi\>/\eps}=\exp\big[\eps^{-1}\!\int k(r)\psi(r)\,dr\big]$, and its functional derivatives are $\delta G/\delta\psi(r)=\eps^{-1}k(r)\,G$ and $\delta^2 G/\delta\psi(r)\delta\psi(r')=\eps^{-2}k(r)k(r')\,G$. Inserting them into Eq.~\eqref{eq:spde_generator}, every factor of $\eps$ cancels:
\begin{equation}
    \label{eq:hamiltonian_field}
    \boxed{\;H[\psi,k] = \int\! dr\; k(r)\,A[\psi](r)
    + \frac{1}{2}\int\! dr\! \int\! dr'\; k(r)\,Q[\psi](r,r')\,k(r')
    \;=\; \<k, A[\psi]\> + \tfrac12\<k, Q[\psi]\,k\> \;}
\end{equation}
and the Lagrangian is
\begin{equation}
    \label{eq:lagrangian_field}
    L[\psi,\dot\psi] = \tfrac12\big\<\dot\psi - A[\psi],\; Q[\psi]^{-1}\big(\dot\psi - A[\psi]\big)\big\> ,
\end{equation}
with $Q^{-1}$ understood as a pseudo-inverse on the range of $Q$ and $L=+\infty$ outside it, as for degenerate noise in Eq.~\eqref{eq:degenerate_lagrangian}. Eq.~\eqref{eq:hamiltonian_field} is Eq.~\eqref{eq:hamiltonian_diffusive} with $b\to A[\psi]$, $D\to Q[\psi]$, and sums over components replaced by integrals over space. All the model dependence therefore sits in two objects: the deterministic drift $A$ and the noise covariance $Q$. The two cases met most often in practice differ only in $Q$:
\begin{itemize}
    \item \emph{Non-conserved noise} (model-A type, under the classification of \cite{hohenberg_theory_1977}), $Q(r,r')=D[\psi(r)]\,\delta(r-r')$: the field is kicked independently at every point, $\<k,Qk\>=\int D\,k^2\,dr$, and $Q^{-1}$ is a local multiplication by $1/D$.
    \item \emph{Conserved noise} (model-B type), where the noise is the divergence of a random flux, $\sqrt\eps\,\nabla\cdot(\sigma\Lambda)$ with $\Lambda$ a vector of white noises: then $Q=-\nabla\cdot(D\nabla\,\cdot)$ with $D=\sigma^2$, and $\<k,Qk\>=\int D\,|\nabla k|^2\,dr$. Here $Q^{-1}$ is an inverse Laplacian, so the Lagrangian is non-local in space. Moreover, $Q$ annihilates spatially uniform momenta, so only fluctuations that conserve the total mass $\int\psi\,dr$ have finite cost \footnote{Conserved noise requires $\int (\dot\psi - A) dr = 0$; the field's total mass is constant only if the drift conserves it too.}.
\end{itemize}
In Sec.~\ref{sec:amb} we present a concrete example from active field theory, \emph{active model B}, and show that it exhibits an unconventional nonequilibrium phase transition \cite{zakine_unveiling_2024}.

\subsection{Weakly interacting particles: the Dawson--G\"artner Hamiltonian}
\label{sec:dawson_gartner}

Consider $N$ diffusions coupled through a pairwise interaction,
\begin{equation}
    \label{eq:weakly_interacting_particles}
    d X_t^{(i)} = \left( b(X^{(i)}_t) + \frac{1}{N}\sum_{j=1}^N c(X_t^{(i)}, X_t^{(j)}) \right) dt + \sigma(X_t^{(i)})\,dW_t^{(i)},
    \qquad i=1,\dots,N .
\end{equation}
The essential difference from Eq.~\eqref{eq:sde} is that the noise on each particle is \emph{not} small. What is large is the number of particles, and the object that becomes deterministic is the empirical density
\begin{equation}
    \label{eq:empirical_densitiy}
    \rho_N(x) = \frac{1}{N}\sum_{i=1}^N \delta(x-X^{(i)}) ,
\end{equation}
which converges as $N\to\infty$ to the solution of the McKean--Vlasov equation. The small parameter is $\eps=1/N$ and the state space is now a space of probability densities.

Similar to Eq.~(\ref{eq:sde_generator}), the $N$-particle generator is
\begin{equation}
    \label{eq:weakly_interacting_particles_generator}
    \mathcal{L}_N = \sum_{i=1}^N\left[b(X^{(i)}) + \frac{1}{N}\sum_{j=1}^N c(X^{(i)}, X^{(j)})\right]\cdot\nabla_i
    + \frac{1}{2} \sum_{i=1}^N D(X^{(i)}):(\nabla\nabla)_i ,
\end{equation}
with $D=\sigma\sigma^\top$. We apply it to the test functional $e^{N\<k,\rho_N\>}$, the natural analogue of $e^{\<k,x\>/\eps}$ when the state is a density. Because $\rho_N$ is a sum of delta functions,
\begin{equation}
    \<k,\rho_N\> = \int \rho_N(x)\,k(x)\,dx = \frac{1}{N}\sum_{i=1}^N k(X^{(i)}),
    \qquad\text{so}\qquad
    e^{N\<k,\rho_N\>} = \prod_{i} e^{k(X^{(i)})} ,
\end{equation}
and therefore
\begin{align}
    \nabla_i\, e^{N\<k,\rho_N\>} & = e^{N\<k,\rho_N\>}\,\nabla k(X^{(i)}), \\
    (\nabla\nabla)_i\, e^{N\<k,\rho_N\>} & = e^{N\<k,\rho_N\>}\Big[\nabla\nabla k(X^{(i)}) + \nabla k(X^{(i)})\otimes \nabla k(X^{(i)}) \Big].
\end{align}
Substituting and dividing by $N$, and abbreviating the self-consistent drift felt by particle $i$ as $b_{\rm eff}(X^{(i)}) = b(X^{(i)}) + \frac1N\sum_j c(X^{(i)},X^{(j)})$,
\begin{equation}
    \frac{1}{N}e^{-N \<k, \rho_N\>}\mathcal{L}_N\, e^{N \<k, \rho_N\>}
    = \frac1N\sum_{i} b_{\rm eff}\big(X^{(i)}\big)\cdot\nabla k(X^{(i)})
    + \frac{1}{2N} \sum_{i} D(X^{(i)}) : \Big[ \nabla\nabla k(X^{(i)}) + \nabla k\otimes\nabla k\big|_{X^{(i)}} \Big].
\end{equation}
As $N\to\infty$ the empirical sums $\frac1N\sum_i$ become integrals $\int\rho(x)\,dx$, giving the Hamiltonian on density space:
\begin{equation}
    \label{eq:hamiltonian_weakly_interacting_particles}
    H(\rho, k) = \int\rho(x) \left[ b(x) + \int c(x,y)\rho(y)\,dy \right] \cdot \nabla k(x)\, dx
    + \underbrace{\frac{1}{2} \int \rho(x)\, D(x) : \nabla\nabla k(x)\, dx}_{\text{deterministic diffusion}}
    + \frac{1}{2} \int \rho(x)\,\big|\sigma^\top(x)\nabla k(x)\big|^2\, dx ,
\end{equation}
where we used $D:(\nabla k\otimes\nabla k) = |\sigma^\top\nabla k|^2$. This is the Hamiltonian corresponding to the large-deviation action for the empirical density derived by Dawson and G\"artner \cite{dawson_large_1987}.

Compared to Eq.~\eqref{eq:hamiltonian_diffusive}, there is one ``deterministic diffusion'' term in Eq.~\eqref{eq:hamiltonian_weakly_interacting_particles}. This term comes from the deterministic evolution of $\rho(t,x)$, and corresponds to the diffusion term in the McKean--Vlasov limit. Recall from Sec.~\ref{sec:H_as_scgf} that the part of $H$ linear in $k$ is always the deterministic evolution. Integrating the first two terms of Eq.~\eqref{eq:hamiltonian_weakly_interacting_particles} by parts (once for the drift, twice for the diffusion),
\begin{equation}
    \int\!\rho\, b_{\rm eff}\!\cdot\!\nabla k \;+\; \tfrac12\!\int\!\rho\, D\!:\!\nabla\nabla k
    = \Big\< k,\; -\nabla\cdot(\rho\, b_{\rm eff}) + \tfrac12\nabla\nabla:(\rho D) \Big\>
    = \big\< k,\; \partial_t\rho \big\>_{\text{McKean--Vlasov}}
\end{equation}
with $b_{\rm eff}(x)=b(x)+\int c(x,y)\rho(y)dy$. The two terms linear in $k$ are precisely the right-hand side of the McKean--Vlasov equation, deterministic drift and deterministic diffusive spreading alike. With that, Eq.~\eqref{eq:hamiltonian_weakly_interacting_particles} has exactly the same two-part structure as Eq.~\eqref{eq:hamiltonian_diffusive}. The same structure underlies the macroscopic fluctuation theory of driven diffusive systems \cite{bertini_macroscopic_2015}.

\section{Rare transitions, quasi-potentials, and metastability}
\label{sec:quasipotential}

We can now provide complete answers to the three questions posed in the introduction. Throughout this section, the deterministic system $\dot x = b(x)$ is assumed to possess multiple attractors $A_1,\dots,A_M$.

Equation~\eqref{eq:ldp_path} gives the least action $V_T(x_a,x_b)$ needed to go from $x_a$ to $x_b$ in a prescribed time $T$. If the duration is not prescribed, we can ask instead: what is the least action needed to reach $x_b$ from $x_a$, however long the path takes? Minimizing over $T$ defines the Freidlin--Wentzell \emph{quasi-potential} \cite{freidlin_random_2012}:
\begin{equation}
    \label{eq:quasi_potential}
    V(x_a,x_b) := \inf_{T>0} V_T(x_a,x_b)
    = \inf_{T>0}\; \inf_{\substack{\phi(0)=x_a\\ \phi(T)=x_b}} S_T[\phi] .
\end{equation}
For a starting attractor $A$, we also minimize over the starting point: $V_A(x)=\inf_{a\in A}V(a,x)$. This cost should not be confused with the probability of \emph{ever} making the transition, which is 1 due to ergodicity. What is exponentially large at weak noise is the mean waiting time to leave the initial well; its exponent is set by the lowest value of $V_A$ on the basin boundary (Sec.~\ref{sec:escape}).

\emph{Divergence of optimal physical transit times.} If the initial state $x_a$ is a stable fixed point of the deterministic flow $\dot x = b(x)$, the optimal path requires an infinite physical time ($T\to\infty$) to depart from $x_a$. This occurs because the deterministic drift vanishes at $x_a$, and moving away with finite speed incurs unnecessary action penalty; the least improbable escape crawls away arbitrarily slowly near the attractor. Consequently, the infimum over $T$ is not attained at any finite $T$, but in the limit $T\to\infty$. This infinite duration is the fundamental reason why naive fixed-time discretization schemes converge poorly near fixed points, necessitating the time-independent geometric reparameterization reviewed in Sec.~\ref{sec:gmam} \cite{heymann_geometric_2008, vanden-eijnden_geometric_2008}.

\subsection{The quasi-potential and the stationary Hamilton--Jacobi equation}
\label{sec:quasipotential_hje}

Fix the starting attractor and write $V(x)\equiv V_A(x)$ for simplicity. Bellman's principle of optimality says that an optimal path to $x$ is optimal on each of its segments. Fix a short final time step $\delta t$. If the velocity on that step is $v$, the path must first reach $x-v\,\delta t$, so minimizing over $v$ gives
\begin{equation}
    V(x) = \inf_v\big\{V(x - v\,\delta t) + L(x,v)\,\delta t\big\} + o(\delta t).
\end{equation}
Assuming $V$ is differentiable at $x$ and expanding $V(x-v\delta t) = V(x) - \<\nabla V, v\>\delta t + o(\delta t)$, then subtracting $V(x)$, dividing by $\delta t$ and letting $\delta t\to0^+$,
\begin{equation}
    0 = \inf_{v}\big\{ L(x,v) - \<\nabla V(x), v\>\big\} = -\sup_{v}\big\{\<\nabla V(x),v\> - L(x,v)\big\} .
\end{equation}
Comparing with Eq.~\eqref{eq:hamiltonian_legendre}, the conjugate momentum is the gradient of the quasi-potential, $k=\nabla V$, and we obtain the stationary Hamilton--Jacobi equation:
\begin{equation}
    \label{eq:stationary_hje}
    \boxed{\;H\big(x, \nabla V(x)\big) = 0 \;}
\end{equation}

We note that while Eq.~\eqref{eq:stationary_hje} always admits the trivial solution $V=\text{const.}$ (corresponding to $k=0$ along deterministic flows), in general it also admits multiple non-trivial solutions. The Dirichlet boundary condition $V(x) = 0$ on the attractor $A$ is usually insufficient to determine the physically meaningful solution. For example, for the linear restoring drift $dX_t=-X_t\,dt+\sqrt{2\eps}\,dW_t$, the equation reads $-xV'(x)+[V'(x)]^2=0$. Both $V(x)=0$ and $V(x)=x^2/2$ solve it and vanish at the stable point $x=0$. The minimum-action definition, Eq.~\eqref{eq:quasi_potential}, tells us which is physical. Starting at the origin, $X_T$ is Gaussian with variance $\eps(1-e^{-2T})$, so the fixed-time cost is $V_T(0,x)=x^2/[2(1-e^{-2T})]$. Its infimum over $T$, i.e., the quasi-potential, is therefore $x^2/2$. 

Physicists usually meet Eq.~\eqref{eq:stationary_hje} by a different route: inserting the WKB ansatz $p_{\rm ss}(x)\asymp e^{-V_{\rm ss}(x)/\eps}$ into the stationary Fokker--Planck or master equation and keeping the leading order in $\eps$. We give this derivation in Appendix~\ref{appendix:wkb}. The stationary exponent $V_{\rm ss}$ and the cost $V_A$ from a specified attractor thus obey the same Hamilton--Jacobi equation, which is the property used in Sec.~\ref{sec:multiple_attractors}. When several attractors compete, the $W$-graph construction in Sec.~\ref{sec:multiple_attractors} combines their local costs into the stationary landscape.

\subsection{Orthogonal decomposition, the Lyapunov property, and the equilibrium limit}
\label{sec:orthogonal}

For diffusions, suppose the drift can be written as the sum of a gradient force and an orthogonal force:
\begin{equation}
    \label{eq:orthogonal_decomposition}
    b = -\tfrac12 D\,\nabla V + F^\perp,
    \qquad \text{with}\qquad \<F^\perp, \nabla V\> = 0 .
\end{equation}
Then, substituting into $H(x,\nabla V) = \<\nabla V, b\> + \tfrac12\<\nabla V, D\nabla V\>$, the first and third terms cancel and the middle vanishes by orthogonality, yielding $H(x,\nabla V)=0$. Conversely, given a solution $V$ of Eq.~\eqref{eq:stationary_hje}, Eq.~\eqref{eq:orthogonal_decomposition} \emph{defines} $F^\perp$ and the orthogonality follows. The decomposition is thus equivalent to the stationary Hamilton--Jacobi equation, and splits the drift into a dissipative part that descends $V$ and a transverse circulation $F^\perp$ that does not change it. The circulation is the signature of broken detailed balance: $F^\perp\equiv0$ if and only if the system is in equilibrium with respect to $V$\footnote{This is exact only when $D$ is constant or $\nabla\cdot D = 0$. In general, for state-dependent $D(x)$, in the It\^o convention, the exact stationary Fokker--Planck current is $J=(b - \eps/2 \nabla\cdot D - \eps/2 D\nabla\log p_{ss})p_{ss}$, so $F^\perp\equiv0$ only guarantees reversibility at leading order in $\eps$, not exact detailed balance at finite $\eps$.}.

Three consequences immediately follow:

\paragraph*{$V$ is a Lyapunov function of the deterministic dynamics.} Along $\dot x = b(x)$,
\begin{equation}
    \frac{d}{dt}V(x(t)) = \<\nabla V, \dot{x}(t)\> = \<\nabla V, b\> = -\tfrac12\<\nabla V, D\,\nabla V\> \le 0 .
\end{equation}
The quasi-potential therefore decreases monotonically under relaxation and is constant only where $\nabla V$ lies in the kernel of $D$. This justifies the picture of $V$ as a landscape, with the important difference that the deterministic flow does not follow $-\nabla V$ but spirals down it.

\paragraph*{The instanton climbs with the circulation preserved.} The optimal escape path is the zero-energy Hamiltonian trajectory with $k=\nabla V$, i.e.
\begin{equation}
    \label{eq:instanton}
    \dot x = \left.\frac{\partial H}{\partial k}\right|_{k=\nabla V} = b + D\,\nabla V = +\tfrac12 D\,\nabla V + F^\perp ,
\end{equation}
while the relaxation dynamics is $\dot x = b(x) = -\tfrac12 D\nabla V + F^\perp$. The dissipative component is reversed while the circulating force remains the same.

\paragraph*{The equilibrium limit.} For $dX = -\nabla U\,dt + \sqrt{2\eps}\,dW$ we have, in the convention of Eq.~\eqref{eq:sde}, $\sigma=\sqrt2\,\mathbb{I}$ and $D=2\,\mathbb{I}$, so $H = -\<\nabla U,k\> + |k|^2$. Then $H(x,\nabla V)=0$ gives $|\nabla V|^2 = \<\nabla U,\nabla V\>$, solved by $\nabla V=\nabla U$, hence
\begin{equation}
    V = U, \qquad p_{\rm ss}(x) \asymp e^{-U(x)/\eps},
\end{equation}
which is the Boltzmann distribution with $\eps=k_BT$. The circulation vanishes, and the quasi-potential is the energy. Freidlin--Wentzell theory therefore contains equilibrium statistical mechanics as its $F^\perp=0$ special case, which is both reassuring and a useful check on any numerical methods.

\subsection{Minimum action path \textit{vs.} minimum energy path}
\label{sec:map_vs_mep}

These two terms used for ``the transition path'' are not synonyms when the system is driven out of equilibrium. The \emph{minimum energy path} (MEP) is an equilibrium object: the path connecting two minima of a potential $U(x)$ that is everywhere parallel to $\nabla U(x)$, i.e.\ the steepest-descent path from the saddle to each minimum. It is what the string methods compute \cite{e_string_2002, e_minimum_2004}.

The \emph{minimum action path} (MAP), or \emph{instanton}, is the minimizer of the action Eq.~\eqref{eq:quasi_potential}. It can be defined for arbitrary drift and noise, and it is a \emph{directed} path: the path from $A_1$ to $A_2$ does not need to be the reverse of the path from $A_2$ to $A_1$. The relation between them is given by Eq.~\eqref{eq:instanton}. Reversing time in the relaxation flow $\dot x = -\tfrac12 D\nabla V + F^\perp$ produces $\dot x = +\tfrac12 D\nabla V - F^\perp$, whereas the instanton follows $\dot x = +\tfrac12 D\nabla V + F^\perp$. Therefore, the uphill instanton is the time reverse of deterministic relaxation if and only if $F^\perp=0$, i.e.\ at equilibrium. In general it is the time reverse of relaxation under the \emph{reversed} flow $\dot x = -\tfrac12 D\nabla V - F^\perp$.

This is the precise content of the familiar statement that out of equilibrium the transition path is not the reversed relaxation path. In Sec.~\ref{sec:examples} and Fig.~\ref{fig:maier_stein}, we illustrate the divergence of these two paths by the Maier--Stein model, with $\beta$ tuning $F^\perp$ away from zero.

\subsection{Escape times and exit locations}
\label{sec:escape}

The headline result of Freidlin--Wentzell theory is the generalization of the Arrhenius law. Let $D$ be a domain whose interior contains a single attractor $x_\ast$, with $\tau_D$ the first exit time. Then for \emph{any} starting point $x$ in the basin,
\begin{equation}
    \label{eq:mean_exit_time}
    \lim_{\eps\to0}\; \eps \log \EE_x[\tau_D] \;=\; \bar V, \qquad \bar V := \min_{y\in\partial D} V_{x_\ast}(y) ,
\end{equation}
that is, the mean escape time is $\EE\tau \asymp e^{\bar V/\eps}$ with $\bar V$ the lowest quasi-potential point on the boundary. Moreover, the exit \emph{location} concentrates on the minimizer: as $\eps\to0$, the probability that the system leaves anywhere other than an arbitrarily small neighborhood of the cheapest boundary point goes to zero.

The independence of Eq.~\eqref{eq:mean_exit_time} from the starting point $x$ is initially surprising. It holds because relaxation to the attractor is fast, i.e., a time of order $1$, compared with escape, which takes a time of order $e^{\bar V/\eps}$. The system forgets where it started long before it escapes, which is also why the escape process is asymptotically exponentially distributed and why the effective Markov chain of Sec.~\ref{sec:multiple_attractors} exists.

\subsection{Several attractors: $W$-graphs and relative stability}
\label{sec:multiple_attractors}

\begin{figure}[tbp]
    \centering
    \includegraphics[width=0.9\textwidth, keepaspectratio]{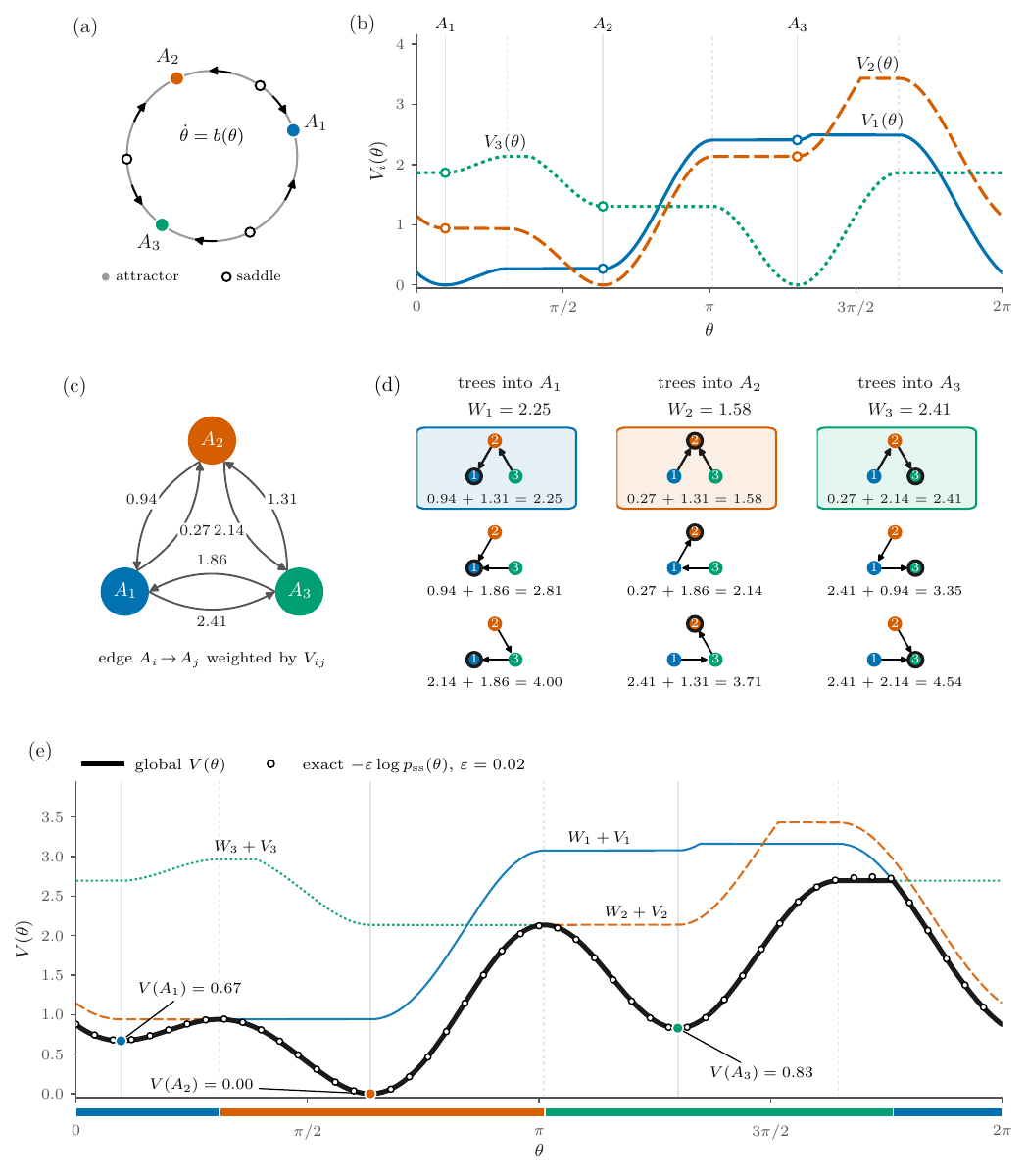}
    \caption{Assembling the global quasi-potential with the $W$-graph construction, for a one-dimensional system on a circle, $d\theta=b(\theta)\,dt+\sqrt{\eps D(\theta)}\,dW_t$, with $b(\theta)=0.1+0.5\sin(\theta+1.6)+0.3\sin(2\theta+1.7)-\sin3\theta$ and $D(\theta)=1+0.3\cos(\theta-0.3)$. (a) Three attractors $A_1,A_2,A_3$ separated by three saddles; the net circulation $\int_0^{2\pi}b/D\,d\theta\neq0$ indicates the system is driven out of equilibrium. (b) Local quasi-potentials $V_i(\theta)$; open circles mark the pairwise costs $V_{ij}=V_i(A_j)$. (c) The pairwise costs as a weighted directed graph. (d) All $i$-trees for each root; the smallest one defines $W_i$ as in Eq.~\eqref{eq:W_graph}. (e) Global quasi-potential, Eq.~\eqref{eq:global_quasipotential}: the lower envelope (thick line) of the lifted local quasi-potentials $W_i+V_i$ (thin lines), with the bar indicating the active branch. Open circles: $-\eps\log p_{\rm ss}$ from the exact stationary solution of the Fokker--Planck equation at $\eps=0.02$. In this example, although from the local quasi-potential in (b) $A_3$ has the highest escape barrier, $A_2$ is stochastically stable, because $A_3$ is harder to reach from $A_1$ or $A_2$. Stochastic stability requires an attractor to be both hard to leave and easy to reach, but pairwise barriers measure only the first.}
    \label{fig:w_graph}
\end{figure}

Suppose the deterministic dynamics has $M$ attractors $A_1,\dots,A_M$. Each of them carries its own local quasi-potential,
\begin{equation}
    V_i(x) \equiv V_{A_i}(x) = \inf_{T>0}\;\inf_{\phi(0)\in A_i,\,\phi(T)=x} S_T[\phi],
\end{equation}
and we can define the pairwise quasi-potential $V_{ij} = V_i(A_j)$ which measures the cost of a transition from $A_i$ to $A_j$ (allowing paths through other attractors). In this subsection we provide a method known as the $W$-graph construction to ``seam'' $V_i(x)$ into a single function $V(x)$ such that the stationary distribution satisfies $p_{\rm ss}(x)\asymp e^{-V(x)/\eps}$ \cite{freidlin_random_2012}.

\paragraph*{Coarse-graining to a Markov chain.} Relaxation into an attractor takes a time of order one, whereas leaving it takes a time of order $e^{V_{ij}/\eps}$ (Sec.~\ref{sec:escape}). On the long time scale the system therefore forgets everything except which attractor it currently sits in, and its motion reduces to a Markov chain on the $M$ attractors with transition probabilities\footnote{Strictly, Freidlin and Wentzell assign to the chain the cost $\tilde V_{ij}$ of paths from $A_i$ to $A_j$ that \emph{avoid} every other attractor, since a path through an intermediate attractor $A_k$ is really two transitions. The minimal tree sums in Eq.~\eqref{eq:W_graph} are the same whether $\tilde V_{ij}$ or $V_{ij}$ is used \cite{freidlin_random_2012}, so one may work with the more easily computed $V_{ij}$.}
\begin{equation}
    q_{ij} \asymp e^{-V_{ij}/\eps}, \qquad i\neq j .
\end{equation}
The relative stability of the attractors is then the stationary distribution $\mu_i$ of this chain.

\paragraph*{The tree construction.} The stationary distribution of a finite Markov chain has an exact combinatorial expression, known as the Markov chain tree theorem \cite{gardiner_stochastic_2009}. Let $G(i)$ be the set of \emph{$i$-trees}: spanning trees on the vertices $\{1,\dots,M\}$ in which every vertex other than $i$ has exactly one outgoing edge, and following the edges from any vertex leads to $i$ without closing a loop (see Fig.~\ref{fig:w_graph}(d) for a concrete example). Assign a weight to each spanning tree $\omega(g) = \prod_{(m\to n)\in g} q_{mn}$, i.e., the product of the transition probabilities on its edges. Then the tree theorem states that the stationary distribution $\mu_i \propto \sum_{g\in G(i)}\omega(g)$.

Inserting $q_{mn}\asymp e^{-V_{mn}/\eps}$ gives $\omega(g) \asymp e^{-\sum V_{mn}/\eps}$, and by Laplace's principle, the sum over trees is dominated by its largest element. Therefore we can define:
\begin{equation}
    \label{eq:W_graph}
    W_i = \min_{g\in G(i)} \sum_{(m\to n)\in g} V_{mn} ,
\end{equation}
and the stationary distribution of the Markov chain:
\begin{equation}
    \label{eq:stationary_weights}
    \mu_i \;\asymp\; \exp\!\left[-\frac{W_i - \min_j W_j}{\eps}\right] .
\end{equation}
Here $\min_j W_j$ is a constant so it could be dropped. By convention we keep it so the minimum of the exponent is 0. The attractors with $W_i=\min_j W_j$ are the \emph{stochastically stable} states; all others have exponentially vanishing occupancy as $\eps\to0$. The order of limits matters: Eq.~\eqref{eq:stationary_weights} describes $t\to\infty$ at fixed small $\eps$, i.e.\ after many transitions have taken place. On times short compared with $e^{\min V_{ij}/\eps}$ the system has not yet left its initial attractor.

\paragraph*{The global quasi-potential.} The same reasoning seams the local quasi-potentials together. To leading order, a visit to a point $x$ is a fluctuation launched from one of the attractors. Launches from $A_i$ occur with probability $\mu_i$, and reaching $x$ from there costs $V_i(x)$, so
\begin{equation}
    p_{\rm ss}(x) \;\asymp\; \sum_{i=1}^M \mu_i\, \exp\left[-\frac{1}{\eps}V_i(x)\right] \;\asymp\; \sum_{i=1}^M \exp\left[-\frac{W_i + V_i(x) - \min_j W_j}{\eps}\right].
\end{equation}
Again we can apply Laplace's principle, and the result is the global quasi-potential of Freidlin and Wentzell \cite{freidlin_random_2012},
\begin{equation}
    \label{eq:global_quasipotential}
    \boxed{\; V(x) = \min_{1\le i\le M}\big[\,W_i + V_i(x)\,\big] - \min_{1\le j\le M} W_j ,
    \qquad p_{\rm ss}(x)\asymp e^{-V(x)/\eps} . \;}
\end{equation}
Each local quasi-potential is lifted by the constant $W_i$, and $V$ is the lower envelope of the lifted functions. In Fig.~\ref{fig:w_graph}, we use a minimal example to show the construction step by step. A few remarks are worth stating explicitly.
\begin{enumerate}
    \item \emph{$V$ satisfies the stationary Hamilton--Jacobi equation}, Eq.~\eqref{eq:stationary_hje}, wherever it is differentiable, because each branch $V_i$ satisfies it and adding a constant $W_i$ does not change the gradient. 
    \item \emph{In general $V$ is not differentiable everywhere}. It is common that $V$ develops a kink when two branches meet (see Fig.~\ref{fig:w_graph}(e)), and Eq.~\eqref{eq:stationary_hje} holds only in the viscosity sense \cite{graham_existence_1984, graham_nonequilibrium_1991, graham_weak-noise_1985}.
    \item \emph{The seams need not lie on basin boundaries.} A point just inside the basin of $A_i$ may be reached more easily, in the sense of Eq.~\eqref{eq:global_quasipotential}, by fluctuations that cross the saddle from a neighboring, more heavily occupied attractor and then relax downhill (see Fig.~\ref{fig:w_graph}(e)).
    \item \emph{The equilibrium limit is recovered.} For $b=-\nabla U$ and $D=2\mathbb{I}$, $W_i-W_j=U(A_i)-U(A_j)$, and Eq.~\eqref{eq:global_quasipotential} returns $V=U-\min U$.
\end{enumerate}

\paragraph*{Two attractors.} For $M=2$ the only $1$-tree is the single edge $2\to1$, so $W_1=V_{21}$ and $W_2=V_{12}$, and Eq.~\eqref{eq:stationary_weights} collapses to
\begin{equation}
    \label{eq:two_state_ratio}
    \lim_{\eps\rightarrow0}\eps \log \frac{\mu_1}{\mu_2} = V_{12} - V_{21} .
\end{equation}
Attractor $A_1$ is favored when it is more expensive to leave $A_1$ than to leave $A_2$. Therefore, the larger occupancy of $A$ means $V_{AB}>V_{BA}$. At the moderate noise used in the simulation in Fig.~\ref{fig:consumer_resource}, however, the occupancies also contain prefactors and finite-$\eps$ corrections (Sec.~\ref{sec:fw_limits}), so they indicate the sign of $V_{AB}-V_{BA}$ rather than its value.

\subsection{Beyond leading-order: what Freidlin--Wentzell theory cannot tell}
\label{sec:fw_limits}

\emph{Prefactors.} Freidlin--Wentzell theory is about determining exponents. The escape rate is $\Gamma = C(\eps)\,e^{-\bar V/\eps}$ and the theory delivers $\bar V$ only; $C$ requires the Gaussian fluctuations around the instanton, which means solving a Riccati equation or evaluating a functional determinant along the optimal path \cite{bouchet_path_2022, heller_evaluation_2024, schorlepp_systematic_2025}. For comparison with experiment or simulation at moderate noise this is usually the dominant discrepancy.

\emph{Finite-$\eps$ corrections.} Separately from the prefactor, the exponent itself acquires corrections in $\eps$. The convergence rate to the asymptotic regime can sometimes be very slow, especially when the prefactor approaches a singularity \cite{debiossac_convergence_2024}.

\emph{Degenerate minimizers.} The theory assumes the infimum is attained, essentially uniquely. When a continuous symmetry acts on the problem, e.g., translation invariance of a field theory, or the rotational invariance of an isotropic system, the minimizer comes in a family; thus, zero modes appear, and the numerical algorithm must be modified \cite{schorlepp_symmetries_2023}.

\emph{Non-instantonic transitions.} The instanton dominates asymptotically, but at any finite noise strength a different mechanism with slightly larger action but much larger phase-space volume can dominate the observed statistics. For example, in transition to turbulence in plane Couette flow, the observed rare transition pathway is not the instanton \cite{rolland_rare_2024}. The correct interpretation of Freidlin--Wentzell theory is as a statement about an asymptotic limit, and the entropic competition between a cheap narrow channel and an expensive wide one is precisely a prefactor question.

\subsection{Examples}
\label{sec:examples}

\subsubsection{Maier--Stein model}
\label{sec:maier_stein}

Consider the (slightly modified) Maier--Stein model \cite{maier_escape_1993, maier_scaling_1996}:
\begin{equation}
\label{eq:maier_stein}
\begin{aligned}
    \dot{x}_1 & = x_1 - x_1^3 - (\beta+1) x_1 x_2^2 + \sqrt{2\eps}\,\xi_1(t) \\
    \dot{x}_2 & = -(1 + x_1^2) x_2 + \sqrt{2\eps}\,\xi_2(t)
\end{aligned}
\end{equation}
where $\xi(t)$ is standard Gaussian white noise, and $\beta\geqslant0$ is a control parameter. This system contains two stable fixed points: $(-1,0)$ and $(1,0)$. When $\beta=0$, Eq.~\eqref{eq:maier_stein} is in equilibrium, since the drift can be written as the gradient of a potential function $U(x_1,x_2) = (1+x_1^2)x_2^2/2 + x_1^4/4 - x_1^2/2$.

Maier and Stein introduced this model as the simplest two-dimensional escape problem in which detailed balance can be switched off continuously \cite{maier_escape_1993}. Physically, it describes an overdamped particle in a symmetric double well, $x_1$ being the reaction coordinate and $x_2$ a transverse degree of freedom, subject to a force field that for $\beta\neq0$ is not the gradient of any potential. Due to its simplicity and symmetry, the model has since become the standard benchmark for numerical methods for instanton calculations \cite{e_minimum_2004, heymann_geometric_2008}.

What makes the model instructive is that $\beta$ does not change the positions of the two stable fixed points and the saddle point. What $\beta$ changes is the rotational part of the drift, $\partial_{x_1}b_2-\partial_{x_2}b_1 = 2\beta\,x_1x_2$, which vanishes on both axes and changes sign from one quadrant to the next. Moreover, the $x_1$-axis is invariant under the deterministic flow, and on it the drift reduces to the one-dimensional gradient $x_1-x_1^3$. The straight path along the axis, uphill from $(-1,0)$ to the saddle, therefore has the same action for every $\beta$, namely $U(0,0)-U(-1,0)=1/4$ (recall $V=U$ for $D=2\mathbb{I}$, Sec.~\ref{sec:orthogonal}). The only question is whether this path remains the cheapest one.

\begin{figure}[tbp]
    \includegraphics[width=\textwidth, keepaspectratio]{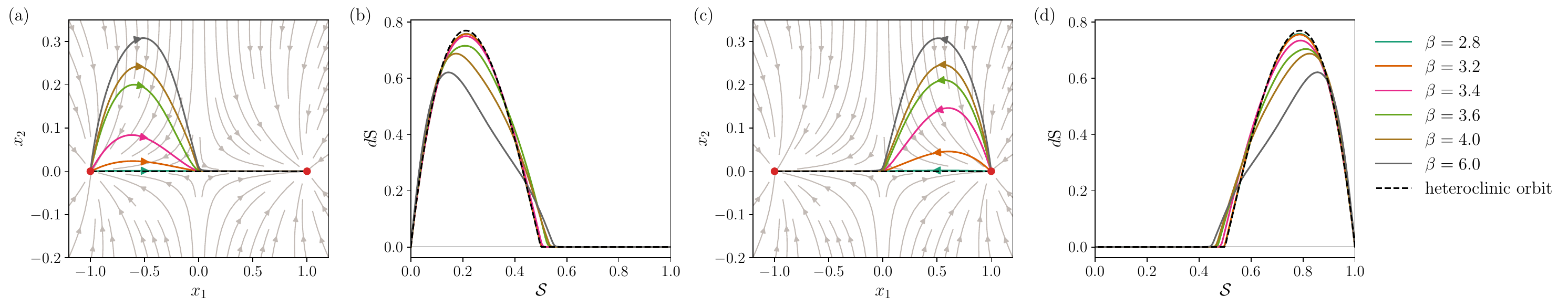}
    \caption{Maier--Stein model, Eq.~\eqref{eq:maier_stein}: forward and backward transition paths diverge as the system is driven out of equilibrium. (a) Forward minimum action paths from $(-1,0)$ to $(1,0)$ for several values of $\beta$, on top of the deterministic flow (grey). (b) Action density along the forward paths as a function of the normalised arclength $s$. (c,d) The same for the backward paths from $(1,0)$ to $(-1,0)$. For $\beta<\beta_c=3$ both paths lie on the $x_1$-axis (dashed line: the heteroclinic orbit); for $\beta>\beta_c$ the uphill branch leaves the axis, so forward and backward paths form a loop. Only the mirror-symmetric branch with $x_2>0$ is shown.}
    \label{fig:maier_stein}
\end{figure}

Fig.~\ref{fig:maier_stein} answers it with MAPs computed numerically (Sec.~\ref{sec:gmam}). For small $\beta$ the minimum action path stays on the axis. For $\beta>\beta_c=3$ the $x_1$-axis path is no longer the MAP. Instead, the forward optimal path from $(-1,0)$ to $(1,0)$ climbs off the axis to the saddle and then relaxes along the axis (panels (b) and (d) show that all the action is accumulated on the uphill branch). The backward path is its mirror image under $x_1\to-x_1$. The forward path and the time reverse of the backward path are therefore different curves, and together they enclose a loop of finite area that widens as $\beta$ increases. This is the general statement of Sec.~\ref{sec:map_vs_mep} in its simplest form: the instanton reverses the dissipative part of the drift but keeps its circulating part $F^\perp$, so uphill and downhill motion follow different routes. The converse is equally instructive. For $0<\beta<3$ the system is out of equilibrium, yet the forward and backward paths coincide on the axis, because the circulation vanishes there. Therefore, \emph{coinciding forward and backward paths are not a test of detailed balance}. Finally, at $\beta=\beta_c$ the minimizer is degenerate in the sense of Sec.~\ref{sec:fw_limits}, and the prefactor of the escape rate shows anomalous scaling with $\eps$, which Maier and Stein analyzed in detail \cite{maier_scaling_1996}.

\subsubsection{Mean-field spin systems}
\label{sec:spins}

Glauber dynamics for a mean-field magnet is a jump process in disguise \cite{bertoin_lectures_1999}, and it provides a strong validation of the formalism because the answer, i.e., the well-known Curie--Weiss free energy \cite{ellis_entropy_2006}, can be derived independently..

Take $N$ Ising spins $s_i=\pm1$ with mean-field energy $-\tfrac{J}{2N}(\sum_i s_i)^2 - h\sum_i s_i$, and let $m=N^{-1}\sum_i s_i$ be the magnetization. A single flip changes $m$ by $\pm2/N$, so in the notation of Sec.~\ref{sec:jump} we have $\Omega=N$, $x=m$, and jump vectors $\nu_\pm=\pm2$. With Glauber rates and $\Phi \equiv \beta(Jm+h)$, the intensive rates are
\begin{equation}
    w_\pm(m) = \underbrace{\frac{1\mp m}{2}}_{\text{fraction available to flip}} \cdot\; \underbrace{\frac{1\pm\tanh\Phi}{2}}_{\text{flip probability}} ,
    \qquad
    H(m,k) = w_+\big(e^{2k}-1\big) + w_-\big(e^{-2k}-1\big).
\end{equation}
The stationary equation $H(m,V')=0$ becomes a quadratic equation with nontrivial root $e^{2V'}=w_-/w_+$, so
\begin{equation}
    V'(m) = \tfrac12\log\frac{w_-}{w_+}
          = \operatorname{artanh}(m) - \beta\,(Jm+h) ,
\end{equation}
using $(1-\tanh\Phi)/(1+\tanh\Phi)=e^{-2\Phi}$. Integrating,
\begin{equation}
    \label{eq:curie_weiss}
    V(m) = \frac{1+m}{2}\log\frac{1+m}{2} + \frac{1-m}{2}\log\frac{1-m}{2}
           - \beta\left(\frac{J}{2}m^2 + h\,m\right) + \text{const}.
\end{equation}
The first two terms are $-s(m)$, minus the entropy per spin of a configuration with magnetization $m$; the last is $\beta$ times the mean-field energy per spin. Eq.~\eqref{eq:curie_weiss} is therefore exactly $\beta f_{\rm CW}(m)$, the Curie--Weiss free energy per spin, and $P(m)\asymp e^{-N\beta f_{\rm CW}(m)}$ is the familiar equilibrium result. \emph{The Freidlin--Wentzell quasi-potential reproduces mean-field thermodynamics}, including the barrier between the two magnetized states below $T_c$ and hence the exponential scale $e^{N\Delta(\beta f)}$ of the switching time.

The value of this example is what happens when the dynamics is driven. Non-reciprocal couplings, self-propulsion, or spins coupled to different baths change $w_\pm$ but not the \emph{form} of the Hamiltonian Eq.~\eqref{eq:hamiltonian_jump_process}. Since there are no gradient structures, the nontrivial root of $H=0$ no longer integrates to anything recognizable as a free energy, so the quasi-potential must be computed. Active spin models of flocking are a concrete case, where the quasi-potential governs the dissipation at the flocking transition \cite{yu_energy_2022}.

\subsubsection{Active model B}
\label{sec:amb}

\begin{figure}[tbp]
    \centering
    \includegraphics[width=\textwidth, keepaspectratio]{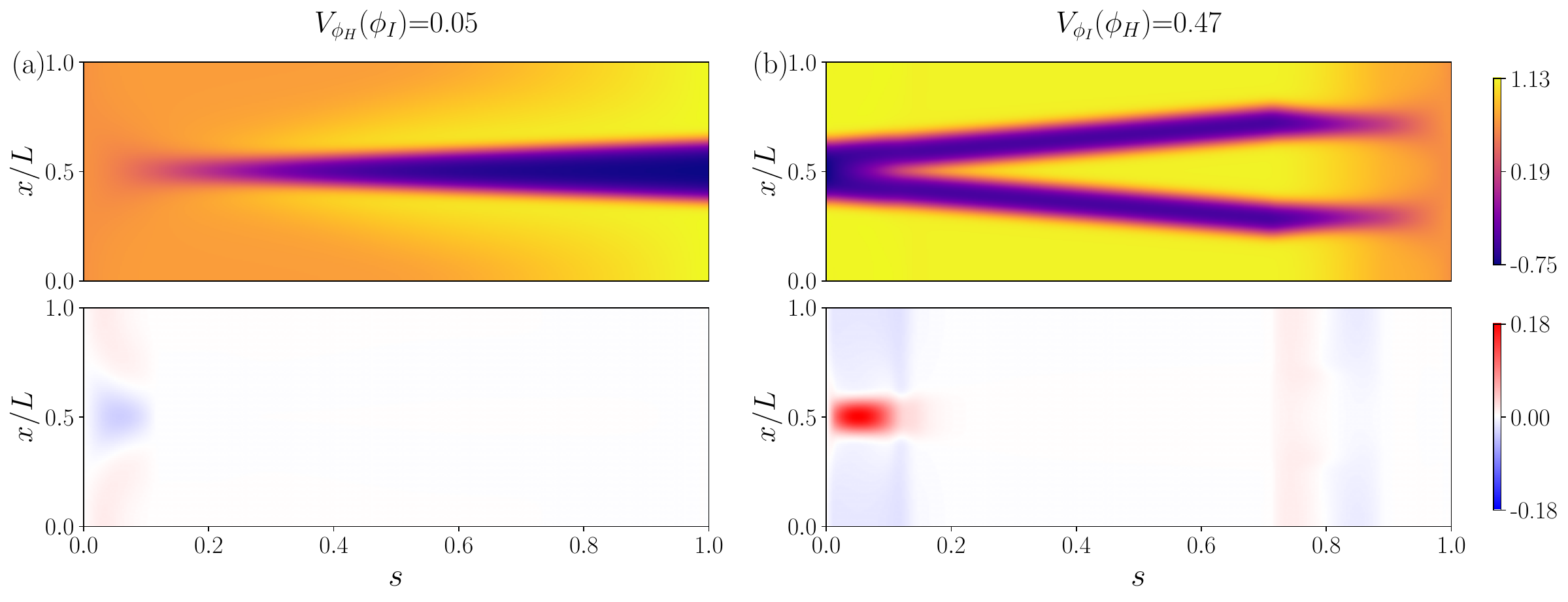}
    \caption{Active model B in one dimension, Eqs.~\eqref{eq:active_model_b}: the transitions between the homogeneous state $\phi_H$ and the phase-separated state $\phi_I$ follow qualitatively different routes in the two directions. (a) Minimum action path from $\phi_H$ to $\phi_I$. Upper panel: the field $\psi(x)$ along the path, as a function of the normalised arclength $s$; lower panel: the conjugate momentum $k(x)$. A single domain of the minority phase is nucleated at the center and then grows deterministically, as in classical nucleation. (b) The same for the reverse path from $\phi_I$ to $\phi_H$. The domain is first split into two, which drift apart and finally dissolve. The momentum is non-zero only during the splitting and in a weak second push near the end, which carries the path across the separatrix. The values above the panels are the two transition costs. Parameters as in Ref.~\cite{zakine_unveiling_2024}: $L=44.7$, $\kappa=\nu=1$, $\lambda=2$, $\phi_0=0.65$, $M=D=1$, with $N_x=256$ grid points.}
    \label{fig:amb}
\end{figure}

As a concrete and physically rich application of Freidlin--Wentzell theory to stochastic field equations, we examine Active Model B, a canonical field-theoretic description of motility-induced phase separation (MIPS) in active matter \cite{wittkowski_scalar_2014, cates_motility-induced_2015, tailleur_active_2022}. In active systems, self-propelled particles undergo macroscopic liquid--gas phase separation into dense and dilute coexisting phases even in the complete absence of attractive interparticle interactions. In the continuum limit, the dynamics govern a conserved scalar density field $\psi(r,t)$ coupled to a noisy diffusive flux:
\begin{align}
    \label{eq:active_model_b}
    \frac{\partial \psi}{\partial t} = -\nabla\cdot J ,
    \qquad
    J = -M[\psi]\,\nabla\mu(\psi) + \sqrt{\eps}\,\sigma[\psi]\,\Lambda ,
\end{align}
with $\Lambda(r,t)$ a vector of standard Gaussian white noises, $M[\psi]$ the mobility, and $\sigma[\psi]$ the noise amplitude. The non-equilibrium chemical potential is given by $\mu(\psi) = f'(\psi) - \kappa\nabla^2\psi + \lambda|\nabla\psi|^2$ and $f(\psi) = a\,\psi^2/2 + b\,\psi^4/4$, with coefficients $a<0<b$ defining a symmetric double-well local free energy density $f(\psi)$, and $\kappa>0$ penalizing interfacial gradients. When activity is present ($\lambda\neq0$), the gradient-squared term $\lambda|\nabla\psi|^2$ cannot be expressed as the functional derivative of any scalar free-energy functional. This non-integrable contribution directly injects energy at interfaces and breaks time-reversal symmetry at the field level, sustaining persistent non-equilibrium probability currents and entropy production \cite{nardini_entropy_2017, markovich_thermodynamics_2021, cates_active_2025}.

The field-theoretic Hamiltonian follows systematically from the generator formula of Sec.~\ref{sec:fields}. The deterministic drift functional is $A[\psi]=\nabla\cdot(M[\psi]\nabla\mu)$. Because the stochastic driving enters as the divergence of a random flux, the noise is conserved, and its spatial covariance operator carries two spatial derivatives:
\begin{equation}
    Q[\psi](r,r') = \EE\big[\nabla_r\!\cdot\!\sigma\Lambda(r)\;\nabla_{r'}\!\cdot\!\sigma\Lambda(r')\big]
    = \nabla_r \cdot \nabla_{r'} \big[ D[\psi(r)]\, \delta(r - r') \big],
    \qquad D=\sigma^2 ,
\end{equation}
yielding $\<k,Qk\>=\int D[\psi]\,|\nabla k|^2\,dr$ after integrating by parts twice. Inserting $A[\psi]$ and $Q[\psi]$ into Eq.~\eqref{eq:hamiltonian_field} and integrating the drift term once by parts yields
\begin{equation}
    \label{eq:hamiltonian_amb}
    H(\psi,k) = \int\! dr\left[-\nabla k\cdot \big(M[\psi]\nabla\mu\big) + \frac{1}{2}\, D[\psi]\,(\nabla k)^2  \right],
\end{equation}
and for constant mobility and noise strength ($M=D=1$), one further integration by parts yields the compact form
\begin{equation}
    \label{eq:hamiltonian_amb_simple}
    H(\psi,k) = \int\! dr\left[(\nabla^2 k)\,\mu(\psi) + \frac{1}{2}\, (\nabla k)^2  \right].
\end{equation}
Here the Laplacian $\nabla^2$ acting on $k$ is the definitive hallmark of conserved dynamics: spatially uniform shifts in momentum ($k = \text{const.}$) yield $H = 0$, guaranteeing that mass-violating fluctuations carry infinite action cost per unit time.

Following Ref.~\cite{zakine_unveiling_2024}, consider the one-dimensional model on a periodic domain of length $L$ with $M=D=1$, $a=-1$, $b=1$ and $\kappa=1$, at mean density $\phi_0=L^{-1}\int\psi\,dx$. In the binodal region the noiseless dynamics has two linearly stable states: the homogeneous state $\phi_H\equiv\phi_0$ and the phase-separated state $\phi_I$. For $\lambda\neq0$ there is no free energy to rank them, so the relative stability must come from Eq.~\eqref{eq:two_state_ratio}. Fig.~\ref{fig:amb} shows the two minimum action paths. The forward path, from $\phi_H$ to $\phi_I$, is very close to classical nucleation: a small nucleus of the minority phase is created at finite cost and then grows deterministically \cite{cates_classical_2023}. 

The reverse transition, from $\phi_I$ to $\phi_H$ [Fig.~\ref{fig:amb}(b)], reveals a completely different, non-equilibrium mechanism. The active force $\lambda|\nabla\psi|^2$ promotes an interfacial instability, causing the central domain to pinch in two under active noise forcing ($k \neq 0$). The resulting subdomains drift apart and pass near an unstable two-domain configuration with two unstable directions, before a second weak momentum push carries the split profile across the separatrix into the homogeneous basin.

\section{Numerical methods}
\label{sec:numerics}

For most physical models, evaluating the quasi-potential and identifying optimal transition pathways requires solving the Hamilton--Jacobi equation Eq.~\eqref{eq:stationary_hje} or minimizing the action Eq.~\eqref{eq:quasi_potential}. Because closed-form analytical solutions exist only in rare symmetric or one-dimensional cases, robust and scalable numerical methods are indispensable. Numerical algorithms in this domain divide into two main paradigms: grid-based Eulerian solvers, such as the Ordered Upwind Method (OUM) for low-dimensional landscapes, and trajectory-based Lagrangian solvers, such as the Geometric Minimum Action Method (gMAM) and its deep neural network variants for high-dimensional diffusions and field theories \cite{lin_data_2022, heller_evaluation_2024, zakine_minimum-action_2023, simonnet_computing_2023, paskal_efficient_2022, kikuchi_ritz_2020, singh_variational_2025}.

\subsection{Ordered upwind method}

The ordered upwind method (OUM) \cite{sethian_ordered_2001, sethian_ordered_2003} is an efficient numerical algorithm designed to solve the stationary Hamilton--Jacobi equation Eq.~\eqref{eq:stationary_hje}. It is the standard grid-based tool for calculating the quasi-potential landscape.

The OUM partitions the grid nodes into three evolving sets: \emph{accepted} (nodes where the viscosity solution $u(x)$ is frozen and finalized), \emph{front} (boundary nodes adjacent to the accepted set carrying tentative values in a min-heap priority queue), and \emph{far} (unvisited nodes initialized to $+\infty$). Starting from the Dirichlet boundary condition $V(A)=0$, the algorithm iteratively accepts the front node with minimal value, updates its unvisited neighbors by solving local characteristic equations, and advances the wavefront until all nodes are accepted. Algorithm~\ref{alg:oum} provides pseudocode for the OUM, as implemented in our accompanying GitHub repository.

Because its computational complexity scales as $O(N \log N)$ (where $N$ is the total number of mesh points), the memory and operation costs scale exponentially with spatial dimension $d$. Consequently, the OUM suffers from the curse of dimensionality and is practical primarily for two- and three-dimensional dynamical systems. Its main advantage is global optimality: as an Eulerian PDE solver, the OUM naturally handles characteristic collisions or caustics, converging unconditionally to the unique global viscosity solution without risk of entrapment in metastable local minima. In practice, for a metastable system with stable attractors $A_1, \dots, A_M$, one executes the OUM once for each attractor $A_i$ subject to the boundary condition $V_i(A_i) = 0$, evaluates the pairwise escape barriers $V_{ij} = V_i(A_j)$, and applies the $W$-graph construction to assemble the global quasi-potential landscape.

\begin{algorithm}[t!]
\caption{Ordered Upwind Method (OUM) \cite{sethian_ordered_2001}}
\label{alg:oum}
\begin{algorithmic}[1]
\Require 2D triangulated mesh $\mathcal{X}$ (max edge length $h$), boundary $\Gamma \subset \mathcal{X}$ (union of mesh edges), boundary data $g: \Gamma \to \mathbb{R}$, speed $f(x, \mathbf{a})$, anisotropy ratio $\Upsilon(x) = F_{\max}(x)/F_{\min}(x)$.
\Ensure Viscosity solution $u(x)$ for all $x \in \mathcal{X}$.

\Statex \textbf{1. Initialization}
\State $\mathcal{A} \gets \Gamma$; \quad $\mathcal{C} \gets \emptyset$; \quad $\mathcal{F} \gets \mathcal{X} \setminus \Gamma$ \Comment{Accepted, Considered (min-heap), Far sets}
\State $u(x) \gets g(x)$ for $x \in \Gamma$, and $u(x) \gets +\infty$ for $x \in \mathcal{F}$
\State $AF \gets \{ [x_i, x_j] \subset \mathcal{A} : \exists \text{ triangle } (x_i, x_j, z) \text{ with } z \notin \mathcal{A} \}$ \Comment{Accepted Front}

\For{$y \in \mathcal{F}$ adjacent to $\mathcal{A}$}
    \State $u(y) \gets \min_{S \in AF,\, d(y, S) \le h\Upsilon(y)} \Call{LocalUpdate}{y, S}$
    \State $\mathcal{F} \gets \mathcal{F} \setminus \{y\}$; \quad $\mathcal{C} \gets \mathcal{C} \cup \{y\}$
\EndFor

\Statex \textbf{2. Main Marching Loop}
\While{$\mathcal{C} \neq \emptyset$}
    \State $x^* \gets \arg\min_{x \in \mathcal{C}} u(x)$; \quad $\mathcal{C} \gets \mathcal{C} \setminus \{x^*\}$; \quad $\mathcal{A} \gets \mathcal{A} \cup \{x^*\}$
    \State Update $AF$ by adding new boundary edges $\Delta AF_{\text{new}}$ incident to $x^*$ and pruning interior edges

    \For{$y \in \mathcal{C}$ and $S \in \Delta AF_{\text{new}}$ with $d(y, S) \le h\Upsilon(y)$} \Comment{Update existing $\mathcal{C}$}
        \State $u(y) \gets \min\big(u(y),\, \Call{LocalUpdate}{y, S}\big)$ and update key in $\mathcal{C}$
    \EndFor

    \For{$y \in \mathcal{F}$ adjacent to $x^*$} \Comment{Promote new neighbors against full $AF$}
        \State $u(y) \gets \min_{S \in AF,\, d(y, S) \le h\Upsilon(y)} \Call{LocalUpdate}{y, S}$
        \State $\mathcal{F} \gets \mathcal{F} \setminus \{y\}$; \quad $\mathcal{C} \gets \mathcal{C} \cup \{y\}$
    \EndFor
\EndWhile

\Statex
\Function{LocalUpdate}{$y, S$} \Comment{1D semi-Lagrangian minimization along $S = [x_i, x_j]$}
    \State Let $x_\tau = (1-\tau)x_i + \tau x_j$ for $\tau \in [0, 1]$, and $\mathbf{a}_\tau = (y - x_\tau)/\|y - x_\tau\|$
    \State \Return $\displaystyle \min_{\tau \in [0, 1]} \left\{ (1-\tau)u(x_i) + \tau u(x_j) + \frac{\|y - x_\tau\|}{f(y, \mathbf{a}_\tau)} \right\}$
\EndFunction
\end{algorithmic}
\end{algorithm}

\subsection{Geometric minimum action method (gMAM)}
\label{sec:gmam}

The OUM works only for low-dimensional systems, so for anything larger we need a method whose cost grows mildly with the dimension. The standard choice is to give up the landscape and compute one transition at a time, by directly minimizing the action over paths that connect two given states $x_a$ and $x_b$. The first such method, the minimum action method (MAM) \cite{e_minimum_2004}, minimizes $S_T[\phi]$ at a fixed $T$. As explained in Sec.~\ref{sec:quasipotential}, the optimal duration of a transition between fixed points is often infinite. A fixed-$T$ discretization therefore wastes most of its points near the endpoints, where the path barely moves, and converges slowly and only as $T\to\infty$. The geometric minimum action method (gMAM) \cite{heymann_geometric_2008, vanden-eijnden_geometric_2008} removes the problem by eliminating time altogether and carries out the minimization on reparametrized curves. Below we first introduce this geometric action, and then present the gradient descent--ascent implementation of Zakine and Vanden-Eijnden \cite{zakine_minimum-action_2023}.

\paragraph*{The geometric action.} Two observations make this possible. First, a minimizer of the free-time problem Eq.~\eqref{eq:quasi_potential} has zero energy, $H(\phi,k)=0$, since otherwise the action could be lowered by changing $T$. Second, on a zero-energy path the Lagrangian reduces to $L=\<k,\dot\phi\>-H=\<k,\dot\phi\>$. Now reparametrize the transition path $\phi$ by a curve $s\in[0,1]$ instead of $t$, write $\varphi(s)=\phi(t(s))$ and $\varphi'=d\varphi/ds$, and let $\lambda(s)=ds/dt\geqslant0$ be the speed at which the curve is traversed, so that $\dot\phi=\lambda\varphi'$. The action becomes $\int\<k,\dot\phi\>dt=\int_0^1\<k,\varphi'\>ds$, which no longer contains the parameterization. Choosing at each point the cheapest admissible momentum gives the geometric action \cite{heymann_geometric_2008}
\begin{equation}
    \label{eq:geometric_action_H}
    V(x_a,x_b) = \inf_{\substack{\varphi(0)=x_a\\ \varphi(1)=x_b}} \hat S[\varphi],
    \qquad
    \hat S[\varphi] = \int_0^1 \<\hat k(s), \varphi'(s)\>\, ds,
    \qquad
    \hat k(s) = \operatorname*{arg\,max}_{k:\;H(\varphi(s),k)=0}\;\<k,\varphi'(s)\>.
\end{equation}
Since $H$ is convex in $k$ (Sec.~\ref{sec:convexity}), the inner maximization is a well-posed convex problem at every $s$. Its Lagrange condition is
\begin{equation}
    \label{eq:gmam_inner}
    H\big(\varphi,\hat k\big) = 0,
    \qquad
    \partial_k H\big(\varphi,\hat k\big) = \lambda\,\varphi', \quad \lambda\geqslant0 ,
\end{equation}
and the multiplier $\lambda$ is exactly the speed $ds/dt$: the second equation is Hamilton's equation $\dot\phi=\partial_kH$ written in the variable $s$. Eq.~\eqref{eq:geometric_action_H} requires only the Hamiltonian, so it applies without change to every model class of Sec.~\ref{sec:dictionary}.

For the diffusive Hamiltonian Eq.~\eqref{eq:hamiltonian_diffusive}, Eq.~\eqref{eq:gmam_inner} can be solved in closed form. Write $|v|_D=\<v,D^{-1}v\>^{1/2}$. Then $\hat k=D^{-1}(\lambda\varphi'-b)$, and the condition $H=0$ gives $\lambda=|b|_D/|\varphi'|_D$, hence
\begin{equation}
    \label{eq:geometric_action_diffusion}
    \hat S[\varphi] = \int_0^1 \Big( |\varphi'|_D\,|b(\varphi)|_D - \<\varphi', D^{-1}b(\varphi)\> \Big)\, ds .
\end{equation}
By the Cauchy--Schwarz inequality the integrand is non-negative, and it vanishes exactly when $\varphi'$ is parallel to $b$, i.e.\ along deterministic trajectories.

\paragraph*{The geometric action as a min-max problem.} The original gMAM of Heymann and Vanden-Eijnden minimizes $\hat S[\varphi]$ directly \cite{heymann_geometric_2008}. At every iteration it computes $\hat k$ exactly at each point of the curve, by an inner Newton loop that solves Eq.~\eqref{eq:gmam_inner}. It then moves the curve along a preconditioned steepest-descent direction of $\hat S$, which involves the second derivatives $\partial_k^2H$ and $\partial_k\partial_xH$. Here we follow instead the formulation of Zakine and Vanden-Eijnden \cite{zakine_minimum-action_2023}, which keeps the momentum as an independent field and needs only the first derivatives of $H$. By the Legendre duality Eq.~\eqref{eq:hamiltonian_legendre}, $S_T[\phi]=\sup_k\int_0^T[\<k,\dot\phi\>-H(\phi,k)]\,dt$. Changing variables from $t$ to $s$ with $dt=ds/\lambda$ turns the minimization over $T$ into a minimization over the function $\lambda(s)$:
\begin{equation}
    \label{eq:gmam_minmax}
    V(x_a,x_b) = \min_{\lambda\geqslant0}\;\min_{\varphi}\;\max_{k}\;\int_0^1 \Big[\<k,\varphi'\> - \lambda^{-1}H(\varphi,k)\Big]\,ds ,
    \qquad \varphi(0)=x_a,\;\varphi(1)=x_b .
\end{equation}
The functional is concave in $k$, because $H$ is convex. Maximizing pointwise over $k$ gives $\partial_kH=\lambda\varphi'$, and minimizing over $\lambda$ gives $H=0$. These are exactly the two conditions of Eq.~\eqref{eq:gmam_inner}, so $\lambda^{-1}$ plays the role of a Lagrange multiplier that enforces zero energy, and the saddle point of Eq.~\eqref{eq:gmam_minmax} is the geometric action.

\paragraph*{Gradient descent--ascent (GDA).} The saddle point is found by descending in $\varphi$ and ascending in $k$ simultaneously, in an artificial time $\tau$, with both gradients preconditioned by $\lambda$:
\begin{equation}
    \label{eq:gda}
    \partial_\tau\varphi = \alpha\big(\lambda\,k' + \partial_xH\big),
    \qquad
    \alpha\,\partial_\tau k = \lambda\,\varphi' - \partial_kH .
\end{equation}
The constant $\alpha>0$ sets the relative rate of the two updates; in machine-learning terminology this is a two-timescale GDA. The fixed points of Eq.~\eqref{eq:gda} are $\lambda\varphi'=\partial_kH$ and $\lambda k'=-\partial_xH$, i.e.\ Eq.~\eqref{eq:hamilton_equations} written in the variable $s$. The speed $\lambda$ is not evolved but updated directly from its optimality condition, which is shown to be the root of $\lambda^2|\varphi'|^2-\lambda\<\partial_kH,\varphi'\>+H=0$ \cite{zakine_minimum-action_2023}:
\begin{equation}
    \label{eq:gda_lambda}
    \lambda = \frac{\big|\<\partial_kH,\varphi'\>\big| + \sqrt{\max\big\{0,\;\<\partial_kH,\varphi'\>^2 - 4H\,|\varphi'|^2\big\}}}{2\,|\varphi'|^2} .
\end{equation}
The absolute value and the $\max$ keep $\lambda\geqslant0$ during the iteration, when $H=0$ holds only approximately. At convergence $H=0$, and Eq.~\eqref{eq:gda_lambda} reduces to $\lambda=\<\partial_kH,\varphi'\>/|\varphi'|^2$.

\paragraph*{Characteristic variables.} The terms $\lambda k'$ and $\lambda\varphi'$ make Eq.~\eqref{eq:gda} a hyperbolic system. It decouples in the variables $u=\varphi+\alpha k$ and $v=\varphi-\alpha k$:
\begin{equation}
    \label{eq:gda_uv}
    \partial_\tau u = \lambda\,u' + f, \qquad
    \partial_\tau v = -\lambda\,v' + g, \qquad
    f = \alpha\,\partial_xH - \partial_kH, \quad
    g = \alpha\,\partial_xH + \partial_kH .
\end{equation}
As $\tau$ increases, $u$ is transported from $s=1$ towards $s=0$, and $v$ from $s=0$ towards $s=1$. The two boundary conditions $\varphi(0)=x_a$ and $\varphi(1)=x_b$ therefore become one condition on the incoming field at each end,
\begin{equation}
    \label{eq:gda_bc}
    v(s=0) = 2x_a - u(s=0), \qquad u(s=1) = 2x_b - v(s=1),
\end{equation}
which makes the problem well posed. This structure suggests the scheme of \cite{zakine_minimum-action_2023}. Each iteration first updates $u$ by a sweep from $s=1$ to $s=0$, then $v$ by a sweep from $s=0$ to $s=1$. Both sweeps use implicit upwind differences for the transport term and explicit reaction terms $f$, $g$, so each update is an explicit recursion along the curve with cost $O(Md)$. Finally, the points are redistributed so that $|\varphi'|$ is constant in $s$, i.e.\ $s$ is the normalised arclength, exactly as in the string method \cite{e_string_2002}. The complete procedure is summarized in Alg.~\ref{alg:gmam}.

\begin{algorithm}[t!]
\caption{Geometric MAM via GDA \cite{zakine_minimum-action_2023}}
\label{alg:gmam}
\begin{algorithmic}[1]
\Require Hamiltonian $H(x,k)$ and derivatives $\partial_x H, \partial_k H$; stable fixed points $x_a, x_b$ ($\partial_k H(x_{a,b},0)=0$); segments $M$ ($\Delta s = 1/M$); step $\Delta\tau > 0$; timescale ratio $\alpha > 0$; tolerance $\delta > 0$.
\Ensure Discrete minimum-action path $\{\varphi_i\}_{i=0}^M$, momenta $\{k_i\}_{i=0}^M$, speeds $\{\lambda_i\}_{i=0}^M$, and $\hat S \approx V_{x_a}(x_b)$.

\Statex \textbf{1. Initialization}
\State Choose initial path $\{\varphi_i\}_{i=0}^M$ with $\varphi_0=x_a$, $\varphi_M=x_b$, and constant $|\varphi_{i+1}-\varphi_i|$
\State $k_i \gets 0$; \quad $u_i \gets \varphi_i + \alpha k_i$; \quad $v_i \gets \varphi_i - \alpha k_i$ \quad for $i=0,\dots,M$

\Statex \textbf{2. Main Loop}
\Repeat
    \State $(\varphi_i, k_i) \gets \big(\tfrac12(u_i+v_i),\, \tfrac{1}{2\alpha}(u_i-v_i)\big)$; \quad $\lambda_i \gets \Call{Speed}{\varphi, k}$; \quad $f_i \gets \alpha\,\partial_x H(\varphi_i,k_i) - \partial_k H(\varphi_i,k_i)$
    \State $\tilde u_M \gets 2x_b - v_M$; \quad $\tilde u_i \gets \dfrac{u_i + \Delta\tau f_i + r_i \tilde u_{i+1}}{1 + r_i}$ with $r_i = \dfrac{\lambda_i \Delta\tau}{\Delta s}$ \quad for $i = M-1,\dots,0$ \Comment{backward sweep}
    \State $(\varphi_i, k_i) \gets \big(\tfrac12(\tilde u_i+v_i),\, \tfrac{1}{2\alpha}(\tilde u_i-v_i)\big)$; \quad $\lambda_i \gets \Call{Speed}{\varphi, k}$; \quad $g_i \gets \alpha\,\partial_x H(\varphi_i,k_i) + \partial_k H(\varphi_i,k_i)$
    \State $\tilde v_0 \gets 2x_a - \tilde u_0$; \quad $\tilde v_i \gets \dfrac{v_i + \Delta\tau g_i + r_i \tilde v_{i-1}}{1 + r_i}$ with $r_i = \dfrac{\lambda_i \Delta\tau}{\Delta s}$ \quad for $i = 1,\dots,M$ \Comment{forward sweep}
    \State $(\bar\varphi_i, \bar k_i) \gets \big(\tfrac12(\tilde u_i+\tilde v_i),\, \tfrac{1}{2\alpha}(\tilde u_i-\tilde v_i)\big)$
    \State Interpolate $\{(\bar\varphi_i, \bar k_i)\}_{i=0}^M$ onto $M+1$ points equally spaced in arclength of $\bar\varphi \to \{(\varphi_i, k_i)\}_{i=0}^M$
    \State $u_i^{\text{new}} \gets \varphi_i + \alpha k_i$; \quad $v_i^{\text{new}} \gets \varphi_i - \alpha k_i$
    \State $e \gets \max_i \big\{ |u_i^{\text{new}} - u_i|,\, |v_i^{\text{new}} - v_i| \big\} / \Delta\tau$; \quad $(u_i, v_i) \gets (u_i^{\text{new}}, v_i^{\text{new}})$
\Until{$e < \delta$}

\Statex \textbf{3. Output}
\State $\hat S \gets \sum_{i=0}^{M-1} \big\langle \tfrac12(k_i + k_{i+1}),\; \varphi_{i+1} - \varphi_i \big\rangle$ \Comment{optional time: $t_i \gets \sum_{j=1}^{i-1} \Delta s / \lambda_j$, since $\lambda_0=\lambda_M=0$}

\Statex
\Function{Speed}{$\varphi, k$} \Comment{Eq.~\eqref{eq:gda_lambda}}
    \State $\varphi_i' \gets (\varphi_{i+1}-\varphi_{i-1})/(2\Delta s)$ \quad (one-sided finite differences at $i=0,M$)
    \State $c_i \gets \langle \partial_k H(\varphi_i, k_i),\, \varphi_i' \rangle$
    \State \Return $\lambda_i = \dfrac{|c_i| + \sqrt{\max\{0,\, c_i^2 - 4H(\varphi_i, k_i)|\varphi_i'|^2\}}}{2|\varphi_i'|^2}$
\EndFunction
\end{algorithmic}
\end{algorithm}

Because the transport terms are treated implicitly, the step $\Delta\tau$ can be taken much larger than an explicit scheme would allow. It is limited mainly by the explicit reaction terms. \cite{zakine_minimum-action_2023} gives a linear stability analysis and shows that only the fixed point matters, so $\Delta\tau$ and $\alpha$ should simply be chosen as large as convergence allows. For stochastic fields, since $\partial_xH$ and $\partial_kH$ contain spatial derivatives, their discretization must be chosen with the same care as for the underlying stochastic PDE.

\paragraph*{Practical remarks.} Several properties of gMAM are worth keeping in mind.
\begin{enumerate}
    \item \emph{Cost.} One iteration costs $O(Md)$ operations and memory $O(Md)$. For a discretized field, $d$ is the number of grid points. The method therefore remains practical for stochastic SPDEs, where the OUM is hopeless \cite{zakine_minimum-action_2023}.
    \item \emph{Locality.} gMAM is a local method in the space of curves, so the minimizer it finds depends on the initial curve. In the Maier--Stein model with $\beta>3$, an initial curve lying on the $x_1$-axis stays on the symmetric, non-optimal axis path, because the symmetry $x_2\to-x_2$ is preserved by the iteration. One should start from several initial curves and compare the actions. Unlike the OUM, gMAM cannot detect that a cheaper path exists elsewhere.
    \item \emph{Two points at a time.} gMAM returns $V(x_a,x_b)$ for one chosen pair of states, not $V(x)$ on a domain. For the $W$-graph construction of Sec.~\ref{sec:multiple_attractors} we therefore need one computation per ordered pair of attractors, i.e.\ $M(M-1)$ runs for $M$ attractors.
    \item \emph{Derivatives.} The scheme needs only $\partial_xH$ and $\partial_kH$, the same derivatives as Hamilton's equations. Non-quadratic Hamiltonians, such as those of jump processes, therefore require no extra work, unlike the inner Newton loop of the original gMAM. For complicated Hamiltonians these derivatives are conveniently generated by automatic differentiation, which is also what makes the gradient-based variant below natural. Zakine and Vanden-Eijnden apply the method to spatially extended systems with both Gaussian and Poissonian (reaction--diffusion) noise \cite{zakine_minimum-action_2023}.
\end{enumerate}

\subsection{Gradient-based optimization and Neural gMAM}
\label{sec:neural_gmam}

Algorithm~\ref{alg:gmam} represents one specialized approach to solving the saddle-point problem Eq.~\eqref{eq:gmam_minmax}: its update rules leverage the characteristic hyperbolic structure of Eq.~\eqref{eq:gda} to construct an unconditionally stable, implicit upwind discretization. A more flexible and versatile alternative is to treat the minimax variational formulation, Eq.~\eqref{eq:gmam_minmax}, directly as an optimization objective and solve it with general-purpose gradient-based algorithms. Discretizing the curve into $M$ segments with step size $\Delta s = 1/M$, the continuous saddle-point problem takes the discrete form
\begin{equation}
    \label{eq:gda_loss}
    \mathcal I[\varphi,k] = \sum_{i=1}^{M-1}\Big[\<k_i,\varphi_i'\> - \lambda_i^{-1}H(\varphi_i,k_i)\Big]\Delta s ,
\end{equation}
which is minimized with respect to the path coordinates $\varphi$ and maximized with respect to the conjugate momenta $k$, with all necessary gradients evaluated via automatic differentiation. This min-max saddle-point architecture is mathematically isomorphic to two foundational frameworks in modern machine learning: generative adversarial networks (GANs) \cite{goodfellow_generative_2020}, where a generator and a discriminator compete in a zero-sum game, and actor--critic architectures in reinforcement learning \cite{konda_actor-critic_1999}. In the large-deviation context, the trajectory $\varphi$ acts as the generator or actor, searching for the least improbable transition pathway connecting $x_a$ and $x_b$, while the conjugate momentum $k$ serves as the critic or discriminator, penalizing unphysical deviations from the zero-energy constraint $H(\varphi, k) = 0$ and matching instantaneous path velocities to Hamiltonian characteristic directions via $\lambda \varphi' = \partial_k H$. Consequently, powerful GDA methods, accelerated by automatic differentiation on GPUs, can be applied to Eq.~\eqref{eq:gda_loss} directly.

\paragraph*{Direct discrete path optimization.} In its most direct formulation, which forms the core of our accompanying Python package (Sec.~\ref{sec:implementation}), the optimization parameters are the discrete nodal values of the trajectory and momentum themselves, $\{\varphi_i, k_i\}_{i=1}^{M-1}$, with the physical endpoints pinned to the fixed attractors ($\varphi_0 = x_a$, $\varphi_M = x_b$). The parameters are updated using alternating Adam steps: several gradient ascent steps on the momentum field $k$ are performed for each gradient descent step on the trajectory $\varphi$. The local speed $\lambda_i$ is evaluated from the exact zero-energy root, Eq.~\eqref{eq:gda_lambda}, at each iteration and treated as detached from the computational graph (held fixed during backward passes) to ensure stable convergence. To maintain a regular parameterization, the curve is periodically re-interpolated onto an equal-arclength parameterization \cite{e_string_2002}. Compared to the characteristic solver in Alg.~\ref{alg:gmam}, direct discrete GDA eliminates the need to analytically derive operator-split update equations: any Hamiltonian that can be coded can be optimized immediately, and adaptive learning-rate optimizers automatically scale step sizes across stiff coordinate directions.

\paragraph*{Deep neural parameterization (Deep gMAM).} When dealing with spatially extended field theories or high-dimensional interacting particle ensembles, discrete nodal representations encounter severe stability and resolution limits. A transformative alternative is to parameterize the trajectory and momentum as continuous functions represented by deep neural networks. This approach, introduced by Simonnet as the \emph{deep gMAM} \cite{simonnet_computing_2023}, was originally formulated for diffusive systems where the Lagrangian and the closed-form geometric action, Eq.~\eqref{eq:geometric_action_diffusion}, are known analytically. The transition trajectory is expressed as
\begin{equation}
    \label{eq:neural_path}
    \varphi_w(s) = (1-s)\,x_a + s\,x_b + s(1-s)\,\mathcal N_w(s),
\end{equation}
where $\mathcal N_w(s)$ is a deep neural network parameterized by trainable weights $w$. Crucially, the ansatz in Eq.~\eqref{eq:neural_path} satisfies the Dirichlet boundary conditions $\varphi_w(0) = x_a$ and $\varphi_w(1) = x_b$ \emph{identically} for any parameter values $w$, obviating the need for artificial boundary loss penalties. The geometric action integral Eq.~\eqref{eq:geometric_action_diffusion} is evaluated via Monte Carlo collocation by sampling points $s_j \sim \mathcal U(0,1)$ along the curve, with the tangent vectors $\varphi_w'(s) = \partial_s \varphi_w(s)$ computed exactly via back-propagation. The network is then trained by minimizing the regularized loss functional
\begin{equation}
    \label{eq:neural_loss}
    \mathcal C(w) = \hat S[\varphi_w] + \gamma_{\rm arc}\Big[\int_0^1|\varphi_w'|^2ds - \Big(\int_0^1|\varphi_w'|\,ds\Big)^2\Big]
\end{equation}
using stochastic gradient descent. The regularization term, weighted by $\gamma_{\rm arc} > 0$, represents the variance of the speed $|\varphi_w'(s)|$ along the path, softly enforcing an equal-arclength parameterization without requiring explicit interpolation or redistribution sweeps. Deep gMAM has demonstrated remarkable success in computing instanton trajectories for nucleation in five-dimensional Allen--Cahn models, extreme events in Burgers turbulence \cite{simonnet_computing_2023}, and the non-equilibrium phase diagram of Active Model B \cite{zakine_unveiling_2024}. Furthermore, when the Lagrangian is unavailable in closed form---as in Markov jump processes---one can parameterize both the trajectory $\varphi_w(s)$ and the momentum $k_v(s)$ via independent neural networks, optimizing the general saddle-point objective Eq.~\eqref{eq:gda_loss} through simultaneous stochastic GDA on $(w,v)$ \cite{simonnet_computing_2023}.

\paragraph*{Practical trade-offs.} Gradient-based and neural formulations offer significant practical advantages: they eliminate the tedious hand-derivation of Euler--Lagrange equations or characteristic schemes, evaluate functional derivatives to machine precision via automatic differentiation, and, in the neural setting, provide mesh-free representations that mitigate the curse of dimensionality for multi-dimensional fields. Moreover, the stochasticity inherent in mini-batch collocation sampling helps optimize trajectories across rugged energy-action landscapes and escape shallow local minima. These benefits, however, involve trade-offs in precision and algorithmic robustness. For low- and moderate-dimensional systems with stiff relaxation dynamics, the characteristic upwind scheme of Alg.~\ref{alg:gmam} converges orders of magnitude faster and achieves higher deterministic accuracy without requiring delicate tuning of learning rates, network architectures, or two-timescale descent--ascent ratios. In practice, the two methodologies are highly complementary: the characteristic gMAM (Alg.~\ref{alg:gmam}) serves as the gold-standard benchmark in low-to-moderate dimensions, while gradient-based and neural architectures extend the reach of Freidlin--Wentzell theory to high-dimensional stochastic fields and complex non-quadratic Hamiltonians.

\subsection{Implementation: from a Hamiltonian to a minimum action path}
\label{sec:implementation}

All the methods of this section need the Hamiltonian and its derivatives. Algorithm~\ref{alg:gmam} needs $\partial_xH$ and $\partial_kH$ at every node of the path. Its gradient-based variant, Eq.~\eqref{eq:gda_loss}, needs gradients of the action with respect to all parameters. To build the local cost of Alg.~\ref{alg:oum} from a general Hamiltonian, one must solve the zero-energy condition for the momentum at each grid point. Deriving these expressions by hand is tedious and error-prone. It is especially so for stochastic fields, where $\partial_xH$ is a functional derivative that involves several integrations by parts (cf.\ Eq.~\eqref{eq:hamiltonian_amb}), and for jump processes, whose Hamiltonians contain exponentials of the momentum.

The accompanying package avoids this step altogether with the automatic differentiation of PyTorch. A model is specified by a single function that returns $H(x,k)$. All derivatives are generated by \texttt{torch.autograd}. Because the values of $H$ at different nodes of the path are independent, a single backward pass through $\sum_iH(\varphi_i,k_i)$ returns $\partial_xH$ and $\partial_kH$ at all nodes simultaneously. The same model object is accepted by the other solvers in the package, including the ordered upwind method, the gradient-based variant of Sec.~\ref{sec:neural_gmam}, and a string method for minimum energy paths and heteroclinic orbits. The $W$-graph construction of Sec.~\ref{sec:multiple_attractors} is provided as well. Analytical derivatives can be supplied when speed matters, and stiff field models can be treated with a semi-spectral variant of Alg.~\ref{alg:gmam}. The source code is available at \href{https://github.com/quark-strange/freidlin-wentzell}{https://github.com/quark-strange/freidlin-wentzell}.

\section{Discussion}
\label{sec:discussion}

Freidlin--Wentzell theory is often perceived as an assortment of disconnected technical results: an action functional for diffusions, a WKB ansatz for master equations, a response-field path integral for field theories, and graph-theoretic constructions for metastability. The central objective of this review has been to show that, from a physical perspective, they represent a single, unified Hamiltonian calculation.

Once the infinitesimal generator $\mathcal L$ of any continuous-time Markov process is identified, the Feng--Kurtz approach, Eq.~\eqref{eq:hamiltonian_generator}, yields the local Hamiltonian $H(x,k)$ in a few straightforward lines. The Hamiltonian immediately gives the Lagrangian $L(x,v)$ and the action $S_T[\phi]$ via Legendre--Fenchel duality, and the action generates the quasi-potential landscape $V(x)$ through the stationary Hamilton--Jacobi equation $H(x,\nabla V)=0$ or through the geometric action $\hat S[\varphi]$. Finally, the combinatorial $W$-graph construction stitches the local quasi-potentials into a global landscape. Along this journey, the three core questions posed in the introduction receive definitive answers:
\begin{enumerate}
    \item The globally most stable attractor is the one minimizing the tree weight $W_i$, which, as demonstrated by the three-attractor system of Fig.~\ref{fig:w_graph}, depends on both escape barriers and transition network connectivity.
    \item Transitions proceed along minimum action instantons, which generically diverge from time-reversed relaxation paths when detailed balance is broken, forming hysteresis loops (Fig.~\ref{fig:maier_stein}).
    \item The mean escape time scales exponentially as $\EE[\tau] \asymp e^{\bar V/\eps}$, governed by the lowest quasi-potential barrier along the basin boundary.
\end{enumerate}

We close by collecting the main limitations of the present framework and the directions in which we expect progress.

\paragraph*{Beyond the exponent.} The theory delivers the exponential scale only. Every comparison with simulation or experiment at finite noise therefore involves the prefactor and finite-$\eps$ corrections, which can sometimes be large \cite{debiossac_convergence_2024}. For nonequilibrium systems, the prefactor requires the Gaussian fluctuations around the instanton, and there are now systematic methods for computing them \cite{bouchet_path_2022, heller_evaluation_2024, schorlepp_systematic_2025}. These methods become singular exactly where the physics is most interesting: when the optimal path bifurcates, as in the Maier--Stein model at $\beta_c$ \cite{maier_scaling_1996}, when a continuous symmetry produces zero modes \cite{schorlepp_symmetries_2023}, or when the quasi-potential develops switching lines and caustics \cite{graham_weak-noise_1985, graham_nonequilibrium_1991}. A practical and robust theory of prefactors for nonequilibrium, high-dimensional systems is still missing. The same applies to the regime in which a path with slightly larger action but a much wider channel dominates at any accessible noise level \cite{rolland_rare_2024}.

\paragraph*{Order of limits and the meaning of $\eps$.} Throughout we assumed one small parameter. Real systems often have several, such as the noise strength, system size, separation of time scales, and, for fields, the lattice spacing. These limits do not need to commute. For stochastic field theories in particular, the weak-noise limit and the continuum limit have to be taken with care, because nonlinear SPDEs with space-time white noise are in general ill-defined in two or more dimensions without renormalization. It is then not obvious which $\eps$ in a coarse-grained model such as active model B corresponds to the physical noise of the underlying particles. Deriving the Hamiltonian of a macroscopic field directly from the microscopic dynamics, as in Sec.~\ref{sec:spins}, is the cleanest way to avoid this ambiguity, but closed coarse-grained Hamiltonians are rare in real cases.

\paragraph*{More general attractors and dynamics.} Most of the machinery is easiest to apply when the attractors are fixed points. Limit cycles, quasi-periodic and chaotic attractors, time-dependent or periodically forced systems, and non-Markovian dynamics with memory are all physically important, e.g.\ in neural activity \cite{nartallo-kaluarachchi_nonequilibrium_2026, hancock_metastability_2025} and biochemical oscillators \cite{lv_energy_2015}. Each of them requires an extension of the framework, either in the definition of the quasi-potential on the attractor or in the Markov property that underlies the generator.

\paragraph*{Computation.} Global quasi-potentials remain limited to two or three dimensions \cite{sethian_ordered_2003, cameron_finding_2012, paskal_efficient_2022}. Path-based methods scale to fields, but they are local and they only connect two given states. In high dimensions, enumerating the relevant attractors and the competing transition channels between them is often harder than computing any single path. For example, in active Model B the initial condition matters a lot, and often we need to sample several initial paths, perform the computation, and identify the correct optimal transition path. Sampling techniques including importance sampling and replica exchange \cite{yan_learning_2022} may be helpful, and new methods such as data-driven \cite{lin_data_2022}, variational \cite{kikuchi_ritz_2020, singh_variational_2025} and neural \cite{simonnet_computing_2023, zakine_unveiling_2024} approaches are promising, but they still lack the error control of classical schemes. Benchmarks with known answers, such as the exactly solvable one-dimensional and gradient cases used in this review, are essential to validate them. 

\paragraph*{Opportunities.} The range of applications is wide and growing. In active matter, the relative stability of phases, and therefore the phase diagram itself, has to be determined by comparing transition costs rather than free energies \cite{zakine_minimum-action_2023, zakine_unveiling_2024, cates_active_2025}. In gene regulatory and biochemical networks, copy numbers are small and the dynamics are jump processes, so the diffusion approximation is exponentially inaccurate for rare switches (Sec.~\ref{sec:jump}) and the Hamiltonian of Eq.~\eqref{eq:hamiltonian_jump_process} is the appropriate starting point \cite{assaf_wkb_2017}. In ecology and climate, the distinction between local and global stability illustrated by the consumer--resource model of Fig.~\ref{fig:consumer_resource} is precisely the question of resilience. Finally, the circulating part $F^\perp$ of the drift and the loop formed by forward and backward paths are geometric signatures of broken detailed balance, which suggests a closer connection between quasi-potentials and stochastic thermodynamics, e.g.\ entropy production and the dissipative cost of maintaining stability \cite{nardini_entropy_2017, fodor_irreversibility_2022, yu_energy_2022}.

The open-source Python package accompanying this review implements the ordered upwind method, the geometric minimum action method and its gradient-based variant, and the $W$-graph construction. We hope that it lowers the barrier for physicists and applied scientists to compute quasi-potentials and transition paths for their own models.

\appendix

\section{Infinitesimal generator of SDE}
\label{appendix:generator}
Here we derive the infinitesimal generator $\mathcal{L}$ for the general It\^o SDE:
\begin{equation}
    dX_t = \mu(X_t, t)\,dt + \sigma(X_t, t)\, dW_t
\end{equation}

For any test function $f(x)\in C^2$, It\^o's formula gives:
\begin{align}
    df = \left(\mu\cdot \nabla f \right) dt + \nabla f\cdot \sigma dW_t + \frac{1}{2} \text{Tr}\left[(\nabla\nabla f) \sigma\sigma^\top\right] dt
\end{align}
Taking the conditional expectation, the martingale noise term $\nabla f\cdot\sigma\,dW_t$ has zero mean and vanishes:
\begin{align}
    \mathbb{E}[df] = \left(\mu\cdot \nabla f \right) dt + \frac{1}{2} \text{Tr}[(\nabla\nabla f) \sigma\sigma^\top] dt .
\end{align}
Defining the diffusion matrix $D(x,t)=\sigma\sigma^\top$, the infinitesimal generator is defined by $\mathcal{L}f = \lim_{\Delta t\to0}\frac{\EE[df]}{\Delta t}$:
\begin{equation}
    \label{eq:appendix_generator}
    \mathcal{L}_t = \mu(x,t)\cdot\nabla + \frac{1}{2}D(x,t):\nabla\nabla
\end{equation}

To avoid confusion with the Laplacian operator, here we use $\nabla\nabla$ to denote the Hessian operator, and `` $:$ '' denotes the Frobenius inner product.

Eq.~\eqref{eq:sde_generator} in the main text follows by specializing Eq.~\eqref{eq:appendix_generator} to the weak-noise scaling of Eq.~\eqref{eq:sde}. There the noise amplitude is $\sqrt\eps\,\sigma$ rather than $\sigma$, so the diffusion matrix appearing in Eq.~\eqref{eq:appendix_generator} is $(\sqrt\eps\sigma)(\sqrt\eps\sigma)^\top = \eps D$ with $D=\sigma\sigma^\top$, which produces the factor $\eps/2$ in Eq.~\eqref{eq:sde_generator}. The same substitution explains why the small parameter survives the derivation of the Hamiltonian in Eq.~\eqref{eq:hamiltonian_diffusive}: the $\eps$ in front of the second-derivative term cancels one of the two factors of $1/\eps$ produced by differentiating $e^{\<k,x\>/\eps}$ twice.

\section{WKB derivation of the Hamilton--Jacobi equation}
\label{appendix:wkb}

For a continuous-time Markov process, the probability density $P^\varepsilon(x, t)$ of arriving at state $x$ at time $t$ satisfies the forward Kolmogorov (Fokker--Planck) equation governed by the forward infinitesimal generator $\mathcal{L}_\varepsilon^\dagger$:
\begin{equation}
    \frac{\partial P^\varepsilon(x, t)}{\partial t} = \mathcal{L}_\varepsilon^\dagger P^\varepsilon(x, t)
\end{equation}

Substituting the WKB ansatz $P^\varepsilon(x, t) = e^{-S(x, t)/\varepsilon}$ yields:
\begin{equation}
    -\frac{1}{\varepsilon}\frac{\partial S(x, t)}{\partial t} e^{-S(x, t)/\varepsilon} = \left( \mathcal{L}_\varepsilon^\dagger e^{-S(x', t)/\varepsilon} \right)\Bigg|_{x'=x}
\end{equation}
Multiplying both sides by $\varepsilon \,e^{S(x, t)/\varepsilon}$ isolates the time derivative:
\begin{equation}
    -\frac{\partial S(x, t)}{\partial t} = \varepsilon e^{S(x, t)/\varepsilon} \left( \mathcal{L}_\varepsilon^\dagger e^{-S(x', t)/\varepsilon} \right)\Bigg|_{x'=x}
\end{equation}

For diffusion processes, the infinitesimal generator is a strictly local differential operator whose action depends solely on the point derivatives evaluated directly at $x$, meaning all terms beyond the linear expansion contribute at order $\mathcal{O}(\varepsilon)$ and vanish under the $\varepsilon \to 0$ limit. Even for jump processes, the transition displacements themselves scale as $\mathcal{O}(\varepsilon)$, restricting the operator's action to an infinitesimal neighborhood where higher-order curvature terms remain asymptotically negligible. Therefore, Taylor expanding $S(x', t)$ at the point $x'=x$, we have:
\begin{equation}
    S(x', t) = S(x, t) + \langle \nabla S(x, t), \, x' - x \rangle + \mathcal{O}(\|x' - x\|^2)
\end{equation}
and substituting into the exponent gives:
\begin{align}
    -\frac{\partial S(x, t)}{\partial t} &= \varepsilon e^{S(x, t)/\varepsilon} \left( \mathcal{L}_\varepsilon^\dagger e^{-\left[ S(x, t) + \langle \nabla S(x, t), \, x' - x \rangle \right]/\varepsilon} \right)\Bigg|_{x'=x} \nonumber = \varepsilon \left( \mathcal{L}_\varepsilon^\dagger e^{-\langle \nabla S(x, t), \, x' - x \rangle / \varepsilon} \right)\Bigg|_{x'=x}
\end{align}

Decomposing the exponent into terms dependent on $x$ and $x'$:
\begin{equation}
    -\frac{1}{\varepsilon}\langle \nabla S(x, t), \, x' - x \rangle = \frac{1}{\varepsilon}\langle \nabla S(x, t), \, x \rangle - \frac{1}{\varepsilon}\langle \nabla S(x, t), \, x' \rangle
\end{equation}
Because the operator $\mathcal{L}_\varepsilon^\dagger$ acts strictly on the spatial variable $x'$, the term $e^{\frac{1}{\varepsilon}\langle \nabla S(x, t), \, x \rangle}$ is a scalar factor that factors out:
\begin{equation}
    -\frac{\partial S(x, t)}{\partial t} = \varepsilon \, e^{\frac{1}{\varepsilon}\langle \nabla S(x, t), \, x \rangle} \left( \mathcal{L}_\varepsilon^\dagger e^{-\frac{1}{\varepsilon}\langle \nabla S(x, t), \, x' \rangle} \right)\Bigg|_{x'=x}
\end{equation}

Let $k = \nabla S(x, t)$ be the canonical momentum. By the adjoint relation $\langle \mathcal{L}_\varepsilon^\dagger u, v \rangle = \langle u, \mathcal{L}_\varepsilon v \rangle$, each spatial derivative $\nabla_{x'}$ in $\mathcal{L}_\varepsilon^\dagger$ transforms under integration by parts with a minus sign relative to $\mathcal{L}_\varepsilon$. Combined with the sign difference in the test functions ($e^{-\frac{1}{\varepsilon}\langle k, x'\rangle}$ versus $e^{+\frac{1}{\varepsilon}\langle k, x'\rangle}$), the leading-order actions coincide:
\begin{equation}
    \lim_{\varepsilon \to 0} \varepsilon \, e^{\frac{1}{\varepsilon}\langle k, x \rangle} \left( \mathcal{L}_\varepsilon^\dagger e^{-\frac{1}{\varepsilon}\langle k, x' \rangle} \right)\Bigg|_{x'=x} = \lim_{\varepsilon \to 0} \varepsilon \, e^{-\frac{1}{\varepsilon}\langle k, x \rangle} \left( \mathcal{L}_\varepsilon e^{\frac{1}{\varepsilon}\langle k, x' \rangle} \right)\Bigg|_{x'=x}
\end{equation}

Taking the limit $\varepsilon \to 0$ and inserting Eq.~\eqref{eq:hamiltonian_def}, we obtain:
\begin{equation}
    -\frac{\partial S(x, t)}{\partial t} = H(x, \nabla S(x, t))
\end{equation}
which is the Hamilton--Jacobi equation governing the action functional:
\begin{equation}
    \frac{\partial S}{\partial t} + H(x, \nabla S) = 0
\end{equation}
It indicates that the Hamiltonian in the HJE corresponding to the rate function $S$ is the Hamiltonian we defined in Eq.~\eqref{eq:hamiltonian_def}.

\bibliography{ref}

\end{document}